\documentclass[12pt, a4paper]{article}

\pdfoutput=1

\usepackage{amsmath}
\usepackage{amsfonts}
\usepackage{amssymb}
\usepackage{bbm}
\usepackage{verbatim}
\usepackage{dsfont}
\usepackage{booktabs}
\usepackage{slashed}
\usepackage{overpic}

\usepackage{epsfig}
\usepackage{color}
\usepackage[table]{xcolor}
\usepackage{graphicx}
\usepackage[]{caption}
\usepackage[listofformat=empty,subrefformat=empty]{subfig}

\usepackage{float}
\usepackage{overpic}

\usepackage{enumerate}
\usepackage{hhline}
\usepackage{multirow}
\usepackage{stackengine}

\stackMath

\usepackage{cite}
\usepackage{xspace}
\usepackage{setspace}
\usepackage{tikz}
\usepackage{footmisc}

\usepackage[pdftitle={},
  pdfauthor={},
  pdfsubject={},
  bookmarksopen, bookmarksnumbered, bookmarksopenlevel=2, colorlinks=false, linkcolor=blue, citecolor=blue, urlcolor=blue]{hyperref}

\begin{document}

\thispagestyle{empty}

\vskip .8 cm
\begin{center}
{ {\bf Towards a Cosmological subsector of Spin Foam Quantum Gravity}}\\[12pt]

\bigskip
\bigskip
{
{{Benjamin Bahr} \footnote{E-mail: benjamin.bahr@desy.de}}, \ {{Sebastian Kl{\"o}ser} \footnote{E-mail: sebastian.kloeser@desy.de}}, \ {{Giovanni Rabuffo} \footnote{E-mail: giovanni.rabuffo@desy.de}}
\bigskip}\\[0pt]
\vspace{0.23cm}
{\it II. Institute for Theoretical Physics\\
University of Hamburg\\
Luruper Chaussee 149\\
22761 Hamburg\\
Germany }\\[20pt]
\bigskip
\end{center}

\begin{abstract}
\noindent
{
\footnotesize
%

We examine the four dimensional path integral for Euclidean quantum gravity in the context of the EPRL-FK spin foam model. The state sum is restricted to certain symmetric configurations which resembles the geometry of a flat homogeneous and isotropic universe. The vertex structure is specially chosen so that a basic concept of expansion and contraction of the lattice universe is allowed.

We compute the asymptotic form of the spin foam state sum in the symmetry restricted setting, and recover a Regge-type action, as well as an explicit form of the Hessian matrix, which captures quantum corrections. We investigate the action in the three cases of vacuum, a cosmological constant, and coupled to dust, and find that in all cases, the corresponding FRW dynamics is recovered in the limit of large lattices. While this work demonstrates a large intersection with computations done in the context of cosmological modelling with Regge Calculus, it is ultimately a setup for treating curved geometries in the renormalization of the EPRL-FK spin foam model.

%
}

\end{abstract}

\newpage
\setcounter{page}{1}
\setcounter{footnote}{0}

{\renewcommand{\baselinestretch}{1.5}
\section{Motivation}
Spin Foam models (SFM) are a promising candidate for a theory of quantum gravity. Built on a path integral formulation, they share many features with the conventional state sum models used to define TQFTs \cite{Barrett:1995mg} and provide a covariant formulation of loop quantum gravity (LQG) \cite{Ashtekar:1995zh, Rovelli:2004tv, Thiemann:2007zz, Rovelli:2010wq, Perez:2012wv}. SFM are defined on a discretization of space-time, which can be regarded as an irregular lattice. Unlike in lattice gauge theory, the lattice does not carry any geometric information. Rather, its geometry  is given by representation-theoretic data distributed among the $d-1$ and $d-2$-dimensional substructures. The sum over all data then realizes a discrete version of the integral over all metrics.


Four-dimensional models for spin foam quantum gravity were first introduced by Barrett and Crane \cite{Barrett:1997gw}, for both Euclidean and Lorentzian signature. Objections were raised about the incorporation of geometrical degrees of freedom and the connection to LQG \cite{Alesci:2007tg}, which is why  several other models emerged in the following years \cite{Engle:2007wy, Freidel:2007py, Baratin:2008du, Baratin:2011hp}. Among these, the model by Engle, Pereira, Livine and Rovelli is one of the most investigated ones. For the value of the Barbero-Immirzi paremter $\gamma$ in the range $\gamma\in(0,1)$, the Riemannian signature version coincides with the model by Freidel and Krasnov, which is why it is called the EPRL-FK model.  While the EPRL model originally was defined only on simplicial complexes, an extension has been proposed in \cite{Kaminski:2009fm}, to incorporate arbitrary polyhedral decompositions of space-time.

This model has received a lot of attention in recent years, since it has many desirable properties. Notably, its large-spin asymptotics is closely connected to discretized general relativity \cite{Barrett:2009gg, Barrett:2009mw, Conrady:2008mk}, in the simplicial case. Also, the model has been recently used to make contact with the cosmological subsector of the theory \cite{Bianchi:2010zs, Rennert:2013pfa}, as well as attempts to compute black hole life-times, which might be connected to observations \cite{Rovelli:2014cta, Christodoulou:2016vny}.

One of the crucial open questions regarding SFM is that of the continuum limit. This is in particular coupled to the problem of renormalization of spin foam models. The renormalization of background-independent theories is generally a non-trivial topic. However, in recent years, there has been a lot of development in this field, in particular on the notion of renormalization and coarse graining in spin foam models \cite{Oeckl:2002ia, Rovelli:2010qx, Bahr:2011aa, Dittrich:2012jq,Bahr:2012qj,  Riello:2013bzw, Bahr:2014qza, Dittrich:2014ala}. Here, a strong connection has been made to the renormalization of tensor networks \cite{Dittrich:2013bza, Dittrich:2013voa, Dittrich:2014mxa, Dittrich:2016tys, Delcamp:2016dqo}, in the context of finite group models and quantum groups. Also, it was observed that the notion of coarse graining is intricately intertwined with the fate of broken diffeomorphism symmetry and the independence under change of discretization in the model \cite{Dittrich:2008pw, Bahr:2009ku, Bahr:2009qc, Bahr:2011uj, Dittrich:2012qb, Banburski:2014cwa}. In the canonical framework this manifests itself with the challenge of constructing an anomaly-free version of the Dirac hypersurface deformation algebra, the constraint algebra for canonical GR \cite{Dittrich:2011vz, Dittrich:2013xwa, Bonzom:2013tna, Dittrich:2014rha}.

In general, one major obstacle towards progress, and also from allowing to use the model to make actual, testable, predictions, is the complexity of SFM, and in particular of the EPRL-FK amplitude. A possible strategy to tackle this issue consists in restricting state sum to certain symmetric configurations. On one hand this approach limits the range of physical systems that can be described by the model, on the other hand it greatly simplifies the expressions of the transition amplitudes. Provided that one can restrict the analysis to a subset of states which dominate the path integral, the sum over such domain would tell us something about the continuum limit of spin foams, expectation values and renormalization group flow of the model.

In the canonical framework, a similar line of thinking has been introduced in \cite{Alesci:2012md}. In the covariant setting, this approach has recently been investigated in \cite{Bahr:2015gxa} in the context of $4d$ Euclidean EPRL-FK Spin Foam model.
Here spacetime is described by a hypercuboidal lattice and the state sum is restricted to coherent intertwiners \cite{Livine:2007vk} that in the large-spin limit resemble a cuboidal geometry. Despite the  drastic reduction of the degrees of freedom the model presents several interesting features. In particular it has been shown that in the semiclassical limit the parameters of the theory tune the restoration of the diffeomorphism symmetry and provide a classification of the dominant states in the path integral.
Under the imposed restrictions, such results open the path to a preliminary analysis of the renormalization properties of spin foams. Recent analysis based on this reduced model have in fact shown numerical evidences of a phase transition in the RG flow \cite{Bahr:2016hwc,Bahr:2017klw}.

A first clear limitation of such a model is the absence of curvature due to vanishing dihedral angles between the cuboidal blocks. In this paper we take the next step along this path by including an elementary form of curvature.
In particular we focus on a discretization in which spacetime is chopped into {\itshape hyperfrusta} i.e., the four dimensional generalization of a truncated regular square pyramid (to which we will in short refer as {\itshape frustum}). The state sum is restricted to coherent intertwiners that in the large-spin limit describe the geometry of a frustum. The emergent curvature is a function of the angle variable that defines the slope of the frustum itself. This extension of degrees of freedom will allow us to go beyond the features of the cuboid model and to forward some cosmological considerations.

The use of the hyperfrustum as the fundamental grain of spacetime is justified by a number of advantages:
\begin{itemize}
	\item A regular hyperfrustum is defined by using just three spins. Consequently, all the formulas that we obtain depend on a quite restricted set of variables. This feature makes the analysis of the model more easy to manage.
	\item The geometry of a hyperfrustum allows a simple and intuitive interpretation as a time-evolving homogeneous and isotropic flat space. Therefore we can use it to model the evolution of a Friedmann universe. Varying the values of the spins one obtains hyperfrusta with different shapes representing spacetimes with different curvature.
	\item The hypercuboidal geometry is found from a particular configuration of the spin variables. In fact a hyperfrustum is a natural extension of the hypercuboidal geometry. Thus we can use the results in \cite{Bahr:2015gxa} as a double check on our computations in the flat spacetime limit.
\item The work in this article is the setup to the extension of the renormalization computations performed in \cite{Bahr:2016hwc, Bahr:2017klw}, in that also states with $4d$ curvature are included in the analysis. We will continue along this line of research in a future article.
\end{itemize}

The paper is organized as follows:
In section \ref{intro} we briefly review the 4d Euclidean EPRL-FK model of quantum gravity. The coherent intertwiners of the theory are constructed in section \ref{symme} and are used to define the edge- and vertex-amplitudes. In section \ref{semicl} we find the complete asymptotic formula for the partition function of the model. We also show that an action appears in the semiclassical limit of the vertex amplitude, which is the generalization of the Regge action to a hypercubic lattice.
In section \ref{modell} we complete the study, by investigating the semiclassical action, and consider its dynamics on larger lattices. Gluing together many hyperfrusta, we set up a tessellation of spacetime. On such a structure we give a qualitative interpretation of the vacuum Friedmann equations in terms of pure geometrical variables (areas, angles, etc..).
Finally, in section \ref{dyn} we study the effective cosmological dynamics starting from the Regge action and deriving the equations of motion. We examine three different cases: the vacuum solution, the universe in presence of a cosmological constant and the coupling of dust particles. A numerical analysis confirms the convergence of our model to the Friedmann universe as the discretization gets refined. Eventually, we show that in the limit of small deficit angles the Regge equations exactly reduce to the standard Friedmann equations, thus suggesting that the restriction of the EPRL-FK model to the symmetric configurations is indeed a viable model for the quantum cosmological subsector of the SFM.

\section{Introduction} \label{intro}

In this paper we analyze the large-spin structure of the 4d Euclidean EPRL-FK spin foam model with Barbero-Immirzi parameter $\gamma<1$ \cite{Engle:2007wy,Freidel:2007py}.
We define the model on a 2-complex $\Gamma$ which is the dual skeleton of our particular discretization of the manifold. The combinatorics of vertices $v$, edges $e$ and faces $f$ in $\Gamma$ is the same of a hypercubic lattice in which all the vertices are eight-valent. In particular each vertex $v$ in $\Gamma$ is dual to a 4d hyperfrustum, and the eight edges meeting at $v$ are dual to the eight 3d hexahedra (two cubes and six pyramidal frusta) which bound the hyperfrustum (see figure \ref{fig:hyperfrustum2}). The faces $f$ of the 2-complex are dual to squares or to regular trapezoids, which in turn form the 2d boundary of cubes and frusta.

\begin{figure}[h]
\centering
	\includegraphics[scale=0.6]{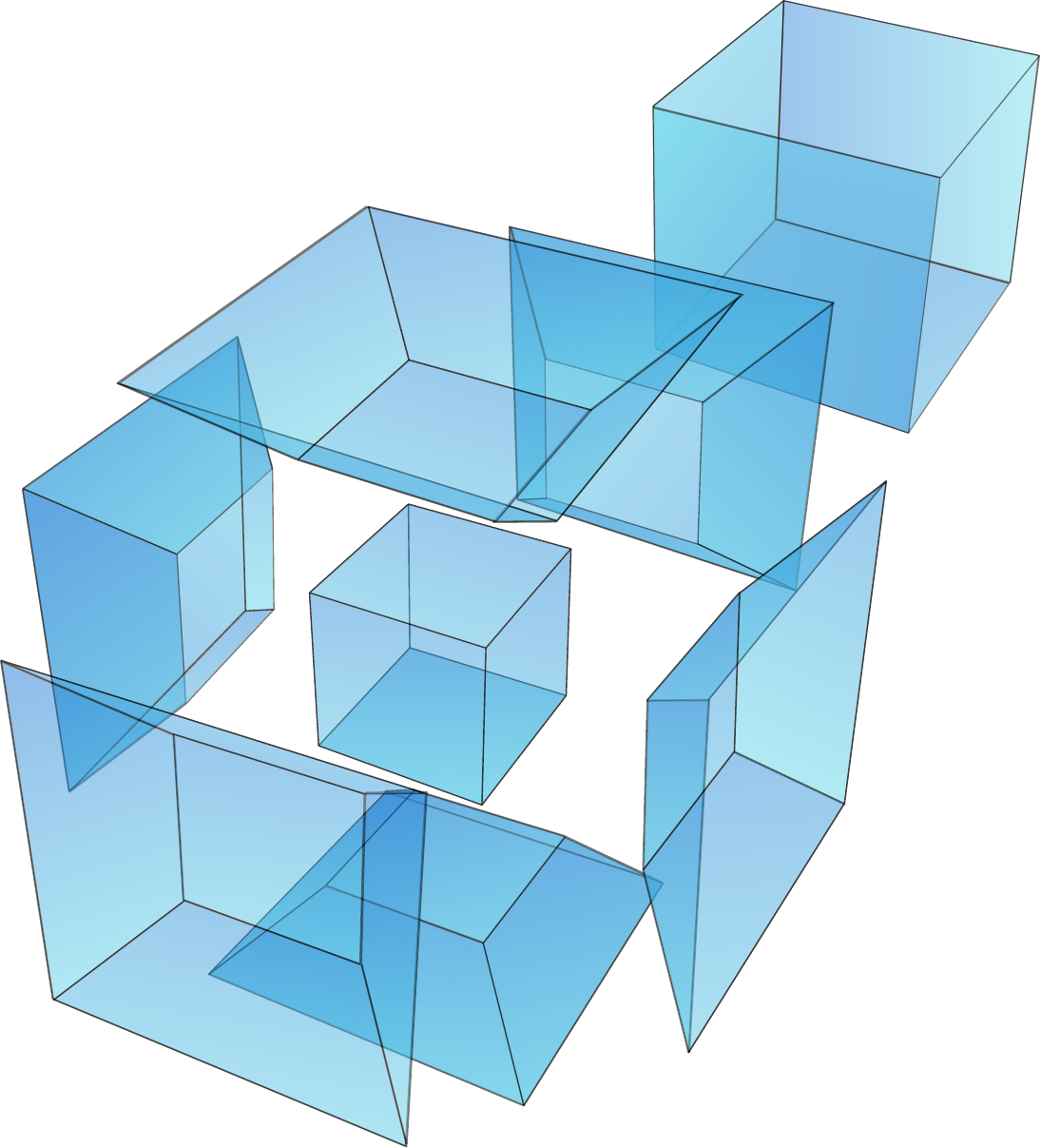}
	\caption{
		\small
		The figure shows the 3d boundary of a hyperfrustum, obtained by unfolding it into six equal frusta and two cubes of different size. This is the analogue, one dimension higher, of the unfolding of a 3d pyramidal frustum into four trapezoids and two squares.}
	\label{fig:hyperfrustum2}
\end{figure}
The hyperfrustum geometry arises by equipping every face $f$ with a spin $j_f$ and every edge $e$ with an intertwiner $\iota_e$.
A 2-complex colored by such specific labeling describes a spacetime configuration in the state sum. Varying the values of the labels in the bulk while keeping fixed the boundary ones amount to consider different `paths' in the path integral. The physical information is deduced from the transition amplitudes between fixed boundary states which belong to the kinematical Hilbert space of LQG.

There exist many equivalent representations of the boundary states in the literature (see for example \cite{Bianchi:2010mw}) and each of them may offer a convenient perspective depending on the kind of problems that one wants to investigate. The usual approach to spin foam cosmology for example relies on the holomorphic representation in which the states have the useful property of being peaked on a specific points of the phase space \cite{Bianchi:2010zs}.
For practical convenience in this paper we make use of the Livine-Speziale coherent state representation \cite{Livine:2007vk}.
The quantum states are thus described by coherent intertwiners which are known to have a simple geometrical interpretation in terms of three dimensional polyhedra. Here we briefly review their construction and extension into the four dimensional context in which they are used to define the transition amplitudes of our spin foam model.

\subsection{Coherent states}

Given the standard eigenstate basis $|j, m\rangle$ of the angular momentum, the maximal weight vector with respect to the  $\hat{e}_3$ direction is $|j, j \rangle$ for any spin $j$.
On such states the dispersion of the angular momentum $\tau_3=\frac{i}{2}\sigma_3$ is minimized and the state becomes a classical polyhedron in the large-$j$ limit. A state fulfilling such properties is called {\itshape coherent state} \cite{Perelomov:1986tf}. Let us take a group element $g\in$ SU(2) and the unit vector $\vec{n}=g \triangleright \hat{e}_3$ defining a direction on the two-sphere $S^2$.
Starting from the maximal weight vector one finds an infinite set of SU(2) coherent states
\begin{equation*}
|j, \vec{n}\rangle = g \triangleright |j, \hat{e}_3 \rangle,
\end{equation*}
for which the angular momentum is minimally spread around $\vec{n}$. Notice that such states are defined up to a U(1) phase, corresponding to a rotation about the $\vec{n}$-direction. Varying  $\vec{n}$ one finds an over-complete set spanning the vector space $V_j$.

Let us consider a set of $N$ coherent states $|j_i, \vec{n}_i\rangle$ such that they satisfy the closure condition $\sum j_i \vec{n}_i=0$.
The basic idea is to associate such states to $N$ faces of area $j_i$ and outward-pointing normals $\vec{n}_i$. A coherent polyhedron is constructed by tensoring them together and imposing the invariance under rotations by SU(2)-group averaging. The associated SU(2) {\itshape coherent intertwiner} reads
\begin{equation}\label{coin}
\iota = \int_{\mathrm{SU(2)}} \mathrm{d} g \ g \triangleright \bigotimes_{i} | j_i , \vec{n}_i\rangle.
\end{equation}
and spans the space $\mathrm{Inv}_{\mathrm{SU(2)}}  \bigotimes_{i} V_{j_i}$ as the vectors $\vec{n}_i$ vary. The SU(2) integration guarantees the invariance under the group action.

The structure just described can be lifted to four dimensions by a {\itshape boosting} procedure which sends
\begin{equation*}
\Phi: \mathrm{Inv}_{\mathrm{SU(2)}} \bigotimes_{i} V_{j_i} \longrightarrow \mathrm{Inv}_{\mathrm{Spin(4)}} \bigotimes_{i} W_i,
\end{equation*}
being $W_i$ a suitable larger space.
In particular the {\itshape boosting map} $\Phi$ consists in the joint action of a map $\beta^{\gamma}_{j_i}$ for each spin $j_i$ such that
\begin{equation}\label{betamap}
\beta^{\gamma}_{j_i}: V_{j_i}  \longrightarrow  W_i,
\end{equation}
and a projector $P$
\begin{equation*}
P: \bigotimes_{i} W_i \longrightarrow  \mathrm{Inv}_{\mathrm{Spin(4)}} \bigotimes_{i} W_i.
\end{equation*}
Given the identification Spin(4)$\simeq$ SU(2)$\times$SU(2)
we can write the vector space $W_i$ as
\begin{equation}
W_i = V_{j_i^+} \otimes V_{j_i^-},
\end{equation}
being $j_i^+$ and $j_i^-$ related to $j_i$ via the Barbero-Immirzi parameter $\gamma$
\begin{equation}\label{jpm}
j_i^{\pm} = \frac{1}{2} |1 \pm \gamma | j_i.
\end{equation}
In the rest of the paper we will focus on the specific case of Barbero-Immirzi parameter $\gamma<1$. In this case the map \eqref{betamap} is defined by embedding the space $V_{j}$ isometrically into the highest weight space of the Clebsh-Gordan decomposition of $V_{j^+} \otimes V_{j^-} $, namely  $V_{j^++j^-}$. Thus we can write
\begin{equation*}
\Phi=P\circ\big(\beta^{\gamma}_{j_1}\otimes \dots \otimes \beta^{\gamma}_{j_N}\big),
\end{equation*}
and the {\itshape boosted coherent intertwiner} reads
\begin{equation}\label{boostedint}
\Phi \iota = \int_{\mathrm{SU(2)} \times \mathrm{SU(2)}} \mathrm{d} g^+  \mathrm{d} g^- \ (g^+ \otimes g^-) \triangleright \bigotimes_{i} | j_i^+ , \vec{n}_i \rangle \otimes | j_i^- , \vec{n}_i\rangle.
\end{equation}
We also refer to $\Phi \iota$ as Spin(4) coherent intertwiner and we are going to use it to build the transition amplitudes in our spin foam model.

\subsection{Transition amplitude}
The transition amplitude is a function of the boundary state, which is defined by assigning a Spin(4) coherent intertwiner $\Phi \iota_n$ to each node and a spin $j_l$ to each link at the boundary of the 2-complex $\Gamma$.
The partition function for the EPRL-FK Euclidean model reads
\begin{equation}\label{generatingfunctional}
Z_{\Gamma}= \sum_{j_{f} \iota_e} \prod_{f} \mathcal{A}_{f} \prod_{e} \mathcal{A}_{e} \prod_{v} \mathcal{A}_{v},
\end{equation}
where the sum is performed over the bulk spins while $\mathcal{A}_{f}$, $\mathcal{A}_{e}$ and $\mathcal{A}_{v}$ are respectively the {\itshape face-},{\itshape edge-} and {\itshape vertex-amplitudes} associated to each element of the 2-complex.

The choice of the face amplitude is not unique and influences the convergence of the state sum \cite{Riello:2013bzw,Bonzom:2013ofa}. We choose the following definition depending on a parameter $\alpha$
\begin{equation}\label{famp}
\mathcal{A}_{f}\equiv [(2 j^{+} + 1)(2 j^{-} + 1)]^{\alpha}.
\end{equation}

The edge amplitude, which is also not unique, is here defined as the normalization of the boosted coherent intertwiner
\begin{equation*}
\mathcal{A}_{e}\equiv \frac{1}{\| \Phi \iota_{e}(j_i) \|^2}.
\end{equation*}
For $\gamma<1$ it factorizes in terms of the SU(2) coherent intertwiners as
\begin{equation}\label{factornorm}
\mathcal{A}_{e}= \frac{1}{\| \iota_{e}(j_i^{+})  \|^2 \| \iota_{e}(j_i^{-})  \|^2}.
\end{equation}

The vertex amplitude is the most important ingredient from which we will recover the Regge action in the semiclassical limit. It is constructed by contracting along links (i.e., boundary edges) the boosted coherent intertwiners at the boundary of each vertex,
\begin{equation}\label{vempl}
\mathcal{A}_v \equiv  \mathrm{tr} \Big( \bigotimes_{e\supset v}\Phi \iota_{e} \Big),
\end{equation}
where we use $\Phi \iota_{e}$ or $(\Phi \iota_{e})^{\dagger}$ depending whether the edge is outgoing or ingoing w.r.t.~the vertex $v$.\\

\section{Reduced Spin Foam Model} \label{symme}

In our model the spin network associated to the boundary of a vertex consists of eight six-valent nodes (see figure \ref{fig:boundhyper}), reflecting the fact that a hyperfrustum is bounded by eight hexahedra: two cubes and six regular pyramidal frusta.
\begin{figure}[h]
	\centering
	\includegraphics[scale=0.8]{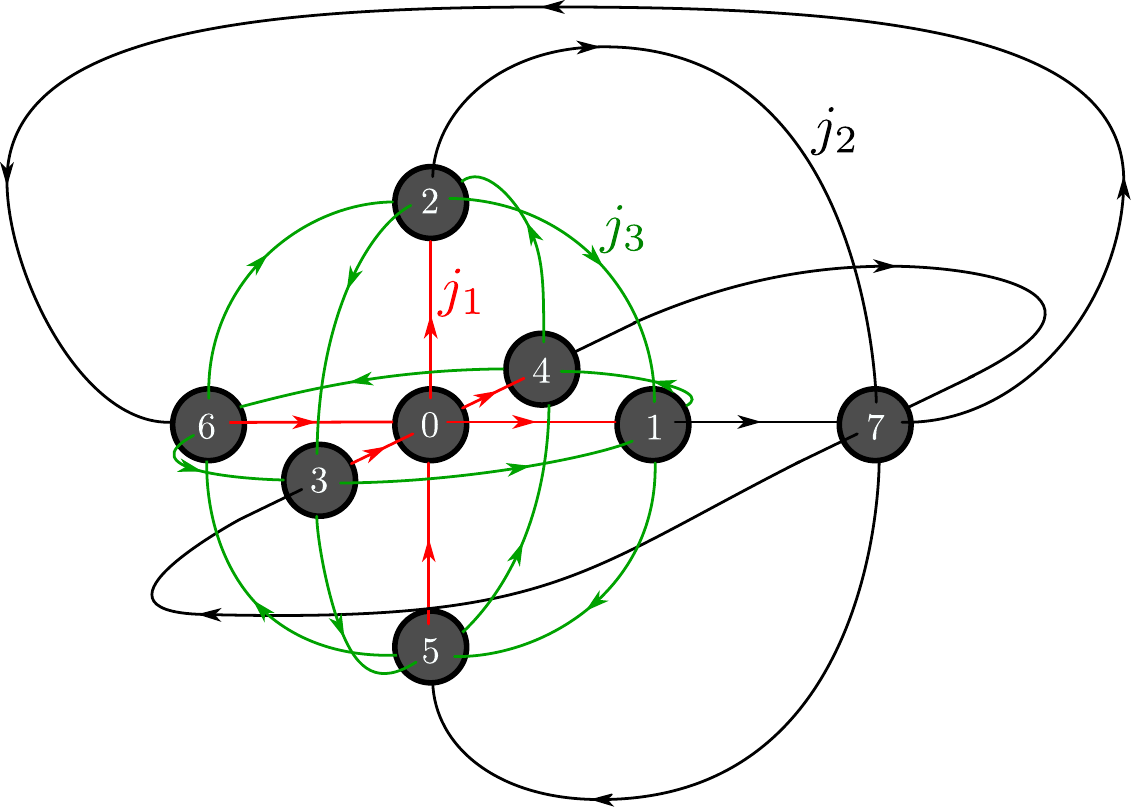}
	\caption{
			\small
			The figure shows the spin network associated to a vertex boundary. This is the dual representation of the 3d boundary in figure \ref{fig:hyperfrustum2}. The three different colors of the links represent three different values of the spins.}
	\label{fig:boundhyper}
\end{figure}

To each node $a=0,\dots,7$  we assign a boosted coherent intertwiner $\Phi \iota_a$ and two SU(2) group elements ($g_{a}^-$, $g_{a}^+$) which account for the group averaging in \eqref{boostedint}.
Each link $ab$ is oriented and is labeled by a spin $j_{ab}$. All the links are automatically endowed with two other spins $j_{ab}^-$ and $j_{ab}^+$ which are related to $j_{ab}$ via the Barbero-Immirzi parameter $\gamma$ as in \eqref{jpm}. The allowed values for $j_{ab}$ are such that $j_{ab}^-$ and $j_{ab}^+$ are half integers. For consistency we also require that $j_{ab}=j_{ba}$.

The coloured spin network just described admits a dual representation in terms of hexahedra $\varepsilon_a$ which are associated to the nodes $a$.
We call $\vec{n}_{ab}\in S^2 \subset \mathbb{R}^3$ the normalized outgoing normal to the face $\Box_{ab}\subset\varepsilon_a$ in the
direction of the neighbouring hexahedron $\varepsilon_b$. The area of $\Box_{ab}$ is given by the spin $j_{ab}$.
The high degree of symmetry chosen ensures that a boundary state is defined by using just three independent values of the spins $j_{ab}$, $\forall a,b,=0\dots7$. We call such values $j_1, j_2, j_3$ and they correspond to the top, bottom and side face areas of any one of the boundary pyramidal frusta represented in figure \ref{fig:hyperfrustum2}.
The previous labeling defines the boundary state and the geometry in our lattice up to a phase factor.\\

\subsection{A note on the boundary data}
Particular attention needs to be paid in defining the initial configuration of the vectors $\vec{n}_{ab}$ at the boundary of a vertex. In fact this choice influences the semiclassical limit of the theory.
In order to clarify this point let us start from the definition of the single vertex amplitude. Usually we build it by forming a closed spin network tensoring together the eight intertwiners $\Phi\iota_a$ at the nodes and joining pairwise the free ends of the links according to the combinatorics (see equation \eqref{vempl} and figure \ref{fig:boundhyper}).
In our case the outcome of this operation depends on the initial choice of the vectors $\vec{n}_{ab}$ which are used to define the coherent intertwiners $\Phi\iota_a$.
For example, embedding the vertex boundary depicted in figure \ref{fig:hyperfrustum2} into a coordinate space and defining the vectors $\vec{n}_{ab}$ accordingly to the oriented axes, one finds out that the asymptotic expression of the vertex amplitude carries a phase factor. Nonetheless, a change of the boundary data can set such phase to zero.
However, at the level of one vertex there are no preferred criteria to chose such initial configuration of the vectors $\vec{n}_{ab}$.
The situation changes if one takes into account the symmetry of a larger structure $\Gamma$ in which many vertices are glued together to form a regular hypercubic lattice.
%
%
For the sake of clarity let us refer to the example in figure \ref{fig:perspect} in which the two vertices $v$,$v^{\prime}\subset \Gamma$ meet along an oriented common edge.
\begin{figure}[h]
\qquad \qquad \quad 	\includegraphics[scale=0.85]{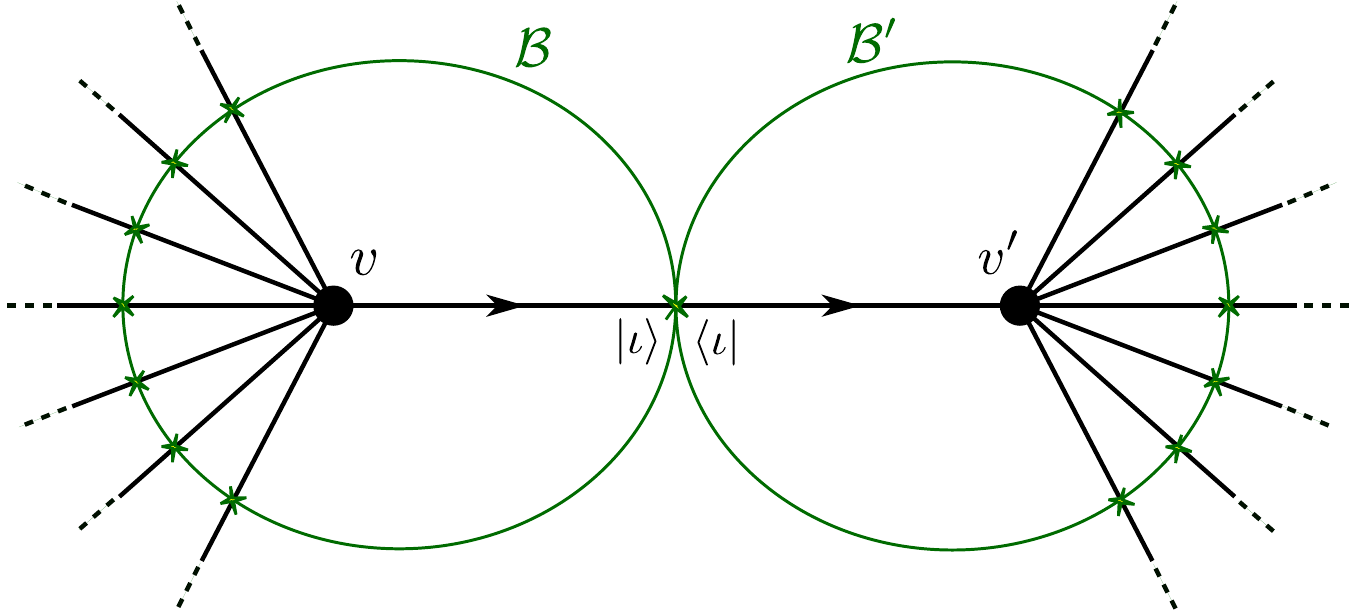}
	\caption{\small The figure shows the gluing of two eight-valent vertices $v$ and $v^{\prime}$ and their respective boundaries (closed lines). In this picture the coherent intertwiners are sitting in the intersections between the boundaries and the edges (straight lines).}
	\label{fig:perspect}
\end{figure}
Here the circles $\mathcal{B}$ and  $\mathcal{B}'$ surrounding the vertices represent their respective boundaries. The intertwiners are placed at the marked intersection points to mean that each of them is associated to an edge $e\subset\Gamma$ and is also an element of a vertex boundary. Let us notice that the intertwiner sitting at the shared edge can be `seen' as an element of $\mathcal{B}$ as well as of $\mathcal{B}'$.
In the bra-ket notation adopted in figure \eqref{fig:perspect} it is denoted by $|\iota\rangle$ or by $\langle \iota|$ depending whether the edge is outgoing or ingoing w.r.t.~the associated vertex.
In a regular lattice the proper gluing of the vertices is such that, given a fixed node $a\subset \mathcal{B}$, the associated intertwiner $\Phi\iota_a$ is contracted to the intertwiner $(\Phi\iota'_{7-a})^\dagger$ in $\mathcal{B}'$. Such (nonlocal) condition must be imposed at all the edges in the lattice. We can however translate this operation in the following (local) constraint on the boundary data of a single vertex
\begin{equation}\label{impos}
(|\vec{n}_{ab}\rangle)^{\dagger} \equiv \langle-\vec{n}_{(7-a) \ b}|, \qquad \qquad  \forall a=4,5,6,7.
\end{equation}

In the dual representation, the example in figure \eqref{fig:perspect} shows two hyperfrusta meeting along a shared hexahedron. Such object lives independently in the boundaries of $\mathcal{B}$ and $\mathcal{B}'$ and it must be identified as the unique hexahedron shared by the two hyperfrusta. In the general case in which the lattice is regular in all the directions, equation \eqref{impos} ensures the proper identification of the boundary hexahedra shared by neighboring hyperfrusta.

With this purpose in mind we can depict the boundary state starting from representing the first four nodes $0,1,2,3$ as in figure \ref{fig:trabound}, and then build the remaining nodes $4,5,6,7$ (dashed lines) respecting the imposition \eqref{impos}.

\begin{figure}[h]
	\includegraphics[scale=0.5]{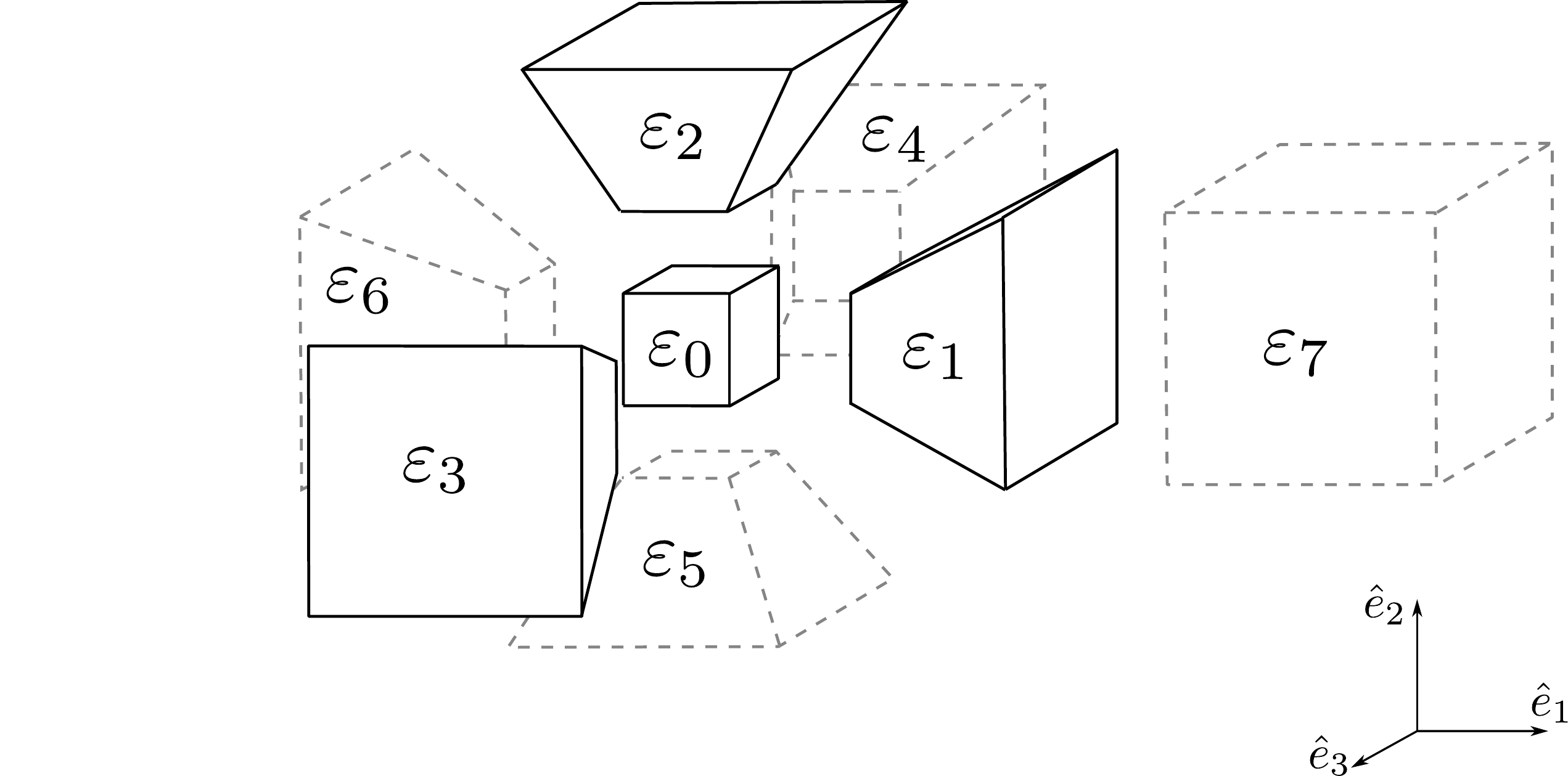}
	
	\caption{	\small The figure shows a specific configuration of the boundary of a vertex. The hexahedra drawn using continuous lines are oriented by relying on the axes $\hat{e}_1,\hat{e}_2,\hat{e}_3$. The remaining hexahedra are defined from the first by applying the condition \eqref{impos}. }
	\label{fig:trabound}
\end{figure}
Remarkably, once the lattice symmetry is taken into account by imposing  \eqref{impos} at the local level, the asymptotic expression of the single vertex amplitude shows no dependence on the choice of phase for the boundary states.

\subsection{Quantum frustum}

The first step towards the definition of the local amplitudes is finding the expressions of the coherent intertwiners. The {\itshape quantum frustum} is a coherent intertwiner that in the large-spin limit describe the geometry of a regular frustum (see figure \ref{fig:frust}). It depends on three spins $j_1,j_2$ and $j_3$ corresponding to its face areas and in the symmetric case $j_1=j_2=j_3=j$ it reduces to a {\itshape quantum cube}.
Thus, this object furnishes a prototype for the description of all the intertwiners appearing in our model.

\begin{figure}[h]
	\centering
	\includegraphics[scale=0.7]{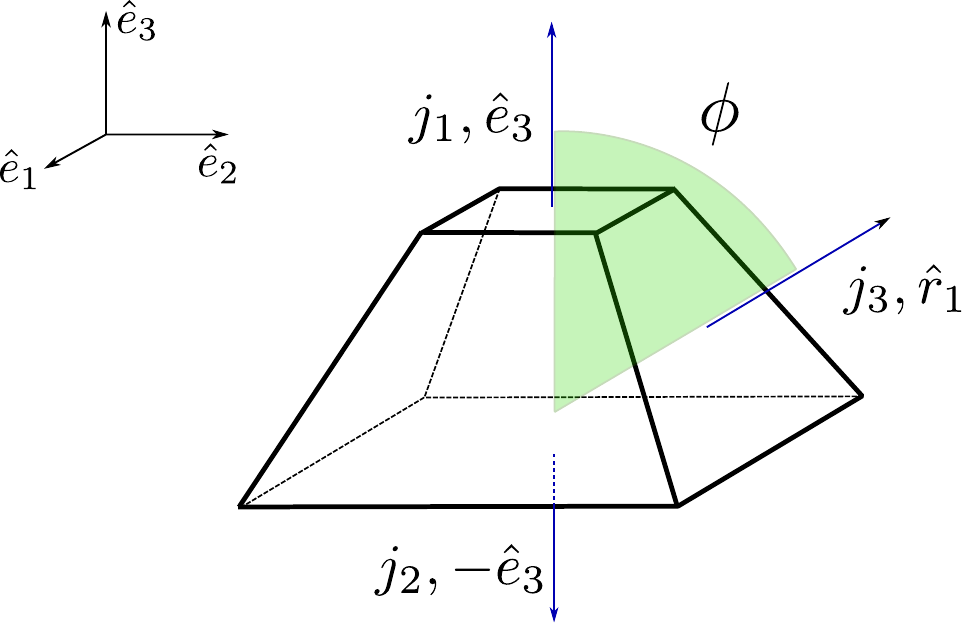}
	\caption{\small The figure shows a {\itshape frustum} i.e., a truncated regular square pyramid}
	\label{fig:frust}
\end{figure}

Following the instructions  given in the previous section, and in particular from \eqref{coin}, we can define a quantum frustum as
\begin{equation}\label{coherentint}
\iota_{j_{1},j_{2},j_{3}} = \int_{SU(2)}  \mathrm{d}  g \ g   \triangleright | j_1 , \hat{e}_3\rangle \otimes g  \triangleright  | j_2 , - \hat{e}_3\rangle \\
\otimes \left( g \triangleright \bigotimes_{l=0}^3 | j_3 , \hat{r}_l\rangle \right),
\end{equation}
where $\hat{r}_l \equiv e^{- i \frac{\pi}{4} l \sigma_3}e^{- i \frac{\phi}{2} \sigma_2}\triangleright\hat{e}_3$ ($l=0,1,2,3$) are the four vectors perpendicular to the side faces of the frustum.
It is possible to express the slope angle $\phi$ of the frustum in terms of the face areas (i.e., the spins) as
\begin{equation}\label{anglespin}
\cos \phi=\frac{j_2-j_1}{4j_3}.
\end{equation}
Using the invariance of the Haar measure to remove one group integration and applying the coherent states property $|j,\vec{n}\rangle = |1/2, \vec{n}\rangle^{\otimes 2j} \equiv  |\vec{n}\rangle^{\otimes 2j}$, the norm of the coherent intertwiner \eqref{coherentint} can be put in the form
\begin{equation}\label{normintertwiner}
\| \iota_{j_{1},j_{2},j_{3}}  \|^2 =  \int_{SU(2)} \mathrm{d} g \ e^{S_e[g]},
\end{equation}
with
\begin{equation}\label{actionnorm}
S_e=2j_1 \ln \langle\hat{e}_3|g|\hat{e}_3\rangle +2j_2 \ln \langle-\hat{e}_3|g|-\hat{e}_3\rangle\\
+2j_3 \sum_{l=0}^{3} \ln  \langle\hat{r}_l|g|\hat{r}_l\rangle .
\end{equation}
In the next section we derive the expression of the edge amplitudes in the large-spin limit starting from the above sample description.
In terms of the coherent states also the vertex amplitude takes a simple and compact form. In particular, for $\gamma<1$ it factorizes as $\mathcal{A}_{v}=\mathcal{A}_{v}^{+} \mathcal{A}_{v}^{-}$ being
\begin{equation}\label{vampl}
\mathcal{A}_{v}^{\pm}=\int_{SU(2)^{8} } \mathrm{d} g_{a} e^{S_{\pm}[g_{a}]} ,
\end{equation}
and
\begin{equation}\label{act}
S_{\pm} [g_{a}]= \frac{|1 \pm \gamma |}{2} \sum_{ab \supset a} 2 j_{ab} \ln \langle- \vec{n}_{ab}|g_{a}^{-1}g_b|\vec{n}_{ab}\rangle \equiv \frac{|1 \pm \gamma |}{2} S_v[g_{a}],
\end{equation}
where we are using the general notation introduced at the beginning of this section to feature the bondary data.

\section{Semiclassical Limit} \label{semicl}
The semiclassical limit of the probability amplitude described by a spin foam model corresponds to the large-spin limit of the partition function \eqref{generatingfunctional}. For the sake of simplicity let us redefine all the spins $j_i \rightarrow \lambda j_i$ so that the asymptotic limit is obtained by sending $\lambda\rightarrow \infty$.
The  limit of the face amplitude is straightforward. From formula \eqref{famp} we obtain
\begin{equation}\label{fam}
\mathcal{A}_{f} \xrightarrow{\lambda \rightarrow \infty} \left[4\lambda^2j^2(1-\gamma^2)\right]^{\alpha}.
\end{equation}

For edge and vertex amplitudes the task is instead not trivial. Notice that the norm of the coherent intertwiner \eqref{normintertwiner} and the vertex amplitude \eqref{vampl} possess a similar form.
To find their large-$\lambda$ limit we will make use of the so called {\itshape extended} stationary phase approximation following reference \cite{Barrett:2009gg}.

\subsection{The extended stationary phase approximation}
The {\itshape extended} stationary phase method provides a tool to compute the asymptotic approximation of oscillatory integrals whose phases are smooth complex valued functions $S$ defined over a closed n-dimensional manifold $X$ and such that $\mathrm{Re} S \leq 0$.
Let us consider the following scalar function
\begin{equation}\label{oscillaintegral}
f(\lambda) = \int_X \mathrm{d} x \ a(x) \ e^{\lambda S(x)},
\end{equation}
being $\lambda$ a positive real parameter and $a(x)$ a smooth complex test function. In the extended stationary phase approximation the asymptotic limit $\lambda \rightarrow \infty$ is dominated by the points $x_0$ such that $\partial_x S|_{x_0}=0$ and $\mathrm{Re} S(x_0) = 0$. These are the stationary and critical points. The leading term in the large-$\lambda$ expansion of \eqref{oscillaintegral} is given by
\begin{equation}\label{extstat}
f(\lambda)\sim\sum_{x_0} \left[a(x_0) \left(\frac{2 \pi}{\lambda}\right)^{n/2} \frac{e^{\lambda S(x_0)}}{\sqrt{\mathrm{det}(-H)}}\right].
\end{equation}
The $n \times n$ Hessian matrix $H$ is given by the second-order partial derivative of $S$ and encodes the informations about the stationary points, which are assumed to be isolated and non-degenerate i.e., $\mathrm{det} H \neq 0$.

Summarizing, in order to compute the asymptotic limit of an oscillatory integral:
\begin{itemize}
	\item we find the critical and stationary points, i.e.~those satisfying $\mathrm{Re} S = 0$ and $dS = 0$.
	\item we compute the Hessian of $S$ in these points and calculate its determinant
	\item we use equation \eqref{extstat} to find the leading term of the large-$\lambda$ limit
\end{itemize}
We are going to use this strategy to compute the large-spin limit of the functions \eqref{normintertwiner} and \eqref{vampl}.

\subsection{The asymptotic norm of the coherent intertwiner}
In order to describe the semiclassical behavior of the edge-amplitude $\mathcal{A}_e$ associated to a quantum frustum we study the large-spin limit of the norm of the coherent intertwiner \eqref{coherentint}. As a first step we look for the critical points of the action $S_e$ in \eqref{actionnorm}.
In our case the manifold carries the structure of a group and the critical points will be $SU(2)$ group elements.
The condition $\mathrm{Re} S_e = 0$ that they have to satisfy can be rephrased in the requirement $|e^{\lambda S_e(x_0)} |=1$. Using the general formula for coherent states
\begin{equation*}
|\langle \vec{n} | \vec{m} \rangle |= \left( \frac{1+\vec{n} \cdot \vec{m}}{2}\right)^{1/2},
\end{equation*}
one finds
\begin{equation*}
\left( \frac{1+\hat{e}_3 \cdot (g\triangleright\hat{e}_3)}{2}\right)^{ j_1}
\left( \frac{1+(-\hat{e}_3) \cdot (g\triangleright(-\hat{e}_3))}{2}\right)^{ j_2}
\prod_l\left( \frac{1+\hat{r}_l \cdot (g\triangleright\hat{r}_l)}{2}\right)^{ j_3} \stackrel{!}{=} 1.
\end{equation*}
Since  the scalar products in the parentheses have real values in the set $[-1,1]$, the above condition is satisfied only for $g = \pm \mathds{1}$. It is easy to check that in these two points the function $S_e$ vanishes. Let us now assign a set of coordinates $x^K, \ K=1,2,3$ to the SU(2) group elements as follows
\begin{equation*}
g \rightarrow g_c e^{\frac{i}{2}x^K \sigma_K}, \qquad g_c=\pm \mathds{1},
\end{equation*}
being $\sigma_K$ the standard Pauli matrices. In these variables $x^K$, the Haar measure is normalized as
\begin{eqnarray}
\frac{1}{(4\pi)^2}\int_{\|x\|<\pi} d^3x\,\left(\frac{\sin(\|x\|/2)}{\|x\|/2}\right)^2\;=\;1.
\end{eqnarray}
This operation allows to perform the partial derivative of the action $S_e$ w.r.t.~the group elements. The first derivative of $S_e$ evaluated in $x=0$ reads
\begin{equation*}
\frac{\partial S_e}{\partial x^K}\Bigg |_{x=0}=i\Bigg(j_1\hat{e}_3^{(K)} -j_2\hat{e}_3^{(K)} +\sum_l j_3\hat{r}_l^{(K)} \Bigg),
\end{equation*}
where we have used the coherent states property $\langle \vec{n} | \sigma_K | \vec{n} \rangle = \vec{n}^{(K)}$ and the expression $\vec{n}^{(K)}$ indicates the $K$-th component of the vector $\vec{n}$ . The above expression is always vanishing since it corresponds to the closure condition.
Thus, we deduce that $g_c= \pm \mathds{1}$ are the critical and stationary points that dominate the asymptotic limit of the norm of the coherent intertwiner.
The components of the Hessian matrix evaluated at the $g_c$ read
\begin{equation*}
H_{KL}=\frac{\partial^2 S_e}{\partial x^L \partial x^K} \Big |_{x=0}= \frac{j_1+j_2}{2}\Big(\hat{e}_3^{(K)}\hat{e}_3^{(L)}-\delta_{KL}\Big) + \sum_{l=0}^{3} \frac{j_3}{2} \Big(\hat{r}_l^{(K)}\hat{r}_l^{(L)}-\delta_{KL}\Big).
\end{equation*}
From the above matrix elements one can derive the determinant of the Hessian
\begin{equation*}
\det(-H)=-\frac{j_3 \sin^2 \phi}{2}\Big(j_1+j_2+2j_3(1+\cos^2 \phi)\Big)^2,
\end{equation*}
where the slope angle $\phi$ is given by \eqref{anglespin}.

Now that we have all the ingredients we can use equation \eqref{extstat} to find the leading term of the norm of the coherent intertwiner \eqref{coherentint} in the large-$\lambda$ expansion.
Inserting the result into equation \eqref{factornorm} we finally obtain the asymptotic limit of the edge amplitude for a quantum frustum
\begin{equation}\label{eamplf}
\mathcal{A}_{e,\mathrm{frustum}}^{j_1,j_2,j_3} \longrightarrow \frac{1}{(4\pi)^4}\Bigg(\frac{\lambda \sqrt{1-\gamma^2}}{8 \pi }\Bigg)^3 j_3 \sin^2 \phi \Big(j_1+j_2+2j_3(1+\cos^2 \phi)\Big)^2.
\end{equation}
From this equation we can easily deduce the large-spin limit of the edge amplitude associated to a quantum cube of side area $j$. By setting $j_1=j_2=j_3\rightarrow j$ we find
\begin{equation}\label{eamplc}
\mathcal{A}_{e,\mathrm{cube}}^j \longrightarrow \frac{1}{16\pi^4}\Bigg(\frac{\lambda \sqrt{1-\gamma^2}}{8 \pi }\Bigg)^3 j^3.
\end{equation}

\subsection{Asymptotics of the vertex-amplitude}
The factorization of the vertex amplitude $\mathcal{A}_v$ for $\gamma<1$ allows us to study its semiclassical limit by focusing on the asymptotic expression of equation \eqref{vampl}. We will make our considerations ignoring the $\pm$ indices and working with the function $S_v$ defined in  \eqref{act}.
The invariance of the Haar measure $\mathrm{d}g$ allows to discard one of the eight integrations by fixing one of the critical points $g_a$. In particular, we choose to fix $g_0=\mathds{1}$. The first condition that the critical points have to satisfy is
\begin{equation}\label{criteq}
|e^{\lambda S_v(x_0)} |=1\Rightarrow g_a \triangleright \vec{n}_{ab}=- g_b\triangleright \vec{n}_{ba}.
\end{equation}
In the geometric picture introduced in the previous section, this condition corresponds to glue the eight boundary hexahedra by properly rotating the vectors $\vec{n}_{ab}$ and  $\vec{n}_{ba}$ so that in the end they will point in relative opposite directions $\forall a$,$b$.
Modulo the symmetry $g_a \rightarrow -g_a$ of the action $S_v$ the critical points equation \eqref{criteq} has two sets of solutions which we list in Table \ref{tab1}.
\begin{table}[h]
\centering
\begin{tabular}{c|c|c}
\rule[-4mm]{0mm}{1cm}
\ & $\Sigma_1$ & $\Sigma_2$ \\
\hline
\rule[-4mm]{0mm}{1cm}
$g_1$ & $\exp(i \frac{\theta}{2} \sigma_1)$ & $\exp(-i \frac{\theta}{2} \sigma_1)$ \\
\rule[-4mm]{0mm}{1cm}
$g_2$ & $\exp(i \frac{\theta}{2} \sigma_2)$ & $\exp(-i \frac{\theta}{2} \sigma_2)$ \\
\rule[-4mm]{0mm}{1cm}
$g_3$ & $\exp(i \frac{\theta}{2} \sigma_3)$ & $\exp(-i \frac{\theta}{2} \sigma_3)$ \\
\rule[-4mm]{0mm}{1cm}
$g_4$ & $\exp(i \frac{\pi-\theta}{2} \sigma_3)$ & $\exp(-i \frac{\pi-\theta}{2} \sigma_3)$ \\
\rule[-4mm]{0mm}{1cm}
$g_5$ & $\exp(i \frac{\pi-\theta}{2} \sigma_2)$ & $\exp(-i \frac{\pi-\theta}{2} \sigma_2)$ \\
\rule[-4mm]{0mm}{1cm}
$g_6$ & $\exp(i \frac{\pi-\theta}{2} \sigma_1)$ & $\exp(-i \frac{\pi-\theta}{2} \sigma_1)$ \\
\rule[-4mm]{0mm}{1cm}
$g_7$ & $\mathds{1}$ & $\mathds{1}$ \\
\end{tabular}
\caption{\small The two sets of critical points which are solutions of equation \eqref{criteq}. We will see that the dihedral angles between the boundary hexahedra are functions of the angle $\theta$ in the exponentials.}
\label{tab1}
\end{table}

The rotation angle $\theta$ can be expressed in terms of the slope angle $\phi$ of the frustum as
\begin{equation}\label{dihedral}
\cos \theta = \frac{1}{\tan \phi}.
\end{equation}
The equation \eqref{dihedral} poses a consistency condition on the allowed values of $\phi$ i.e.,
\begin{equation*}
\frac{\pi}{4}\leq\phi \leq \frac{3\pi}{4}.
\end{equation*}
Using equation \eqref{anglespin} it is easy to check that the allowed values of the spins in our system are
\begin{equation}\label{allowedspins}
-\frac{1}{\sqrt{2}}\leq \frac{j_2-j_1}{4j_3} \leq \frac{1}{\sqrt{2}},
\end{equation}
which correspond to a restriction of the phase space.
The action in the two sets of critical points listed in Table \ref{tab1} reads
\begin{equation*}
	\begin{split}
	S_v(\Sigma_1)&=+6 i (j_1-j_2) \Big(\frac{\pi}{2}-\theta\Big) + 12 i j_3\Big(\frac{\pi}{2}-\arccos\big(\cos^2\theta\big)\Big),\\
	S_v(\Sigma_2)&=-6 i (j_1-j_2) \Big(\frac{\pi}{2}-\theta\Big) - 12 i j_3\Big(\frac{\pi}{2}-\arccos\big(\cos^2\theta\big)\Big).
	\end{split}
\end{equation*}
The Hessian is a $21\times21$ matrix and is constructed with the second derivatives of the action \eqref{act}. Defining the vectors $ \tilde{n}_{ab}\equiv g_a\triangleright\vec{n}_{ab}$, its components evaluated on the critical points are
\begin{equation*}
\begin{split}
	H_{aa,KL}&=\frac{\partial^2 S_v}{\partial x_a^L \partial x_a^K} \Bigg |_{x=0}= \sum_{(ab)\supset a} \frac{j_{ab}}{2} \Bigg(-\delta_{KL}+\tilde{n}_{ab}^{(K)} \tilde{n}_{ab}^{(L)}\Bigg),\\
	H_{ab,KL}&=\frac{\partial^2 S_v}{\partial x_b^L \partial x_a^K} \Bigg |_{x=0}= \frac{j_{ab}}{2}\Bigg(\delta_{KL}-i\epsilon_{KLI}\tilde{n}_{ab}^{(I)}+\tilde{n}_{ab}^{(K)} \tilde{n}_{ab}^{(L)}\Bigg).
\end{split}
\end{equation*}
Using a computer algebra program it is possible to calculate the exact expression of the determinant of the Hessian matrix $D\equiv \det H$, which is a homogeneous function of the spins. The full expression of the determinant relative to the first set $\Sigma_1$ of critical points reads
\begin{equation}\label{determH}
	\begin{split}
	D(j_1,j_2,j_3) = \ &16 \ \lambda^{21} j_1^{3} \ j_2^{3}  \ j_3^{15}  \Big(-1+2\cos^2\phi - iK\Big) \Big(-2+\cos^2\phi +iK\Big)^2\\
	& \Bigg(1 + \cos^2\phi +\frac{j_1+j_2}{2j_3}\Bigg)^3 \Bigg(1+2\cos^2\phi  + \frac{j_1+j_2}{2j_3} - i K\Bigg)^3  \\
	&\Bigg(1+\cos^2\phi + \Big(1-\cos^2\phi \Big)\frac{j_1+j_2}{j_3} + iK \Big(1  -3 \cos^2\phi \Big)\Bigg)^3,
	\end{split}
\end{equation}
with
\begin{equation}\label{K}
K\equiv \sqrt{-\cos 2\phi}=\sqrt{1-2\Big(\frac{j_1-j_2}{4j_3}\Big)^2}.
\end{equation}
Notice that the expression under the square root is always positive for the allowed values \eqref{allowedspins} of the spins, therefore $K$ is a real function with values in the set $[0,1]$. In particular $K=1$ corresponds to the flat cuboid case while $K=0$ corresponds to a degenerate frustum with $\phi=\frac{\pi}{4},\frac{3 \pi}{4}$.
The solution for the set of critical points $\Sigma_2$ is simply given by the complex conjugate of \eqref{determH} which corresponds to send $K\rightarrow -K$.

Computing the leading order \eqref{extstat} for both $\mathcal{A}_{v}^{+}$ and $\mathcal{A}_{v}^{-}$ and taking their product one obtains the leading order of the vertex amplitude $\mathcal{A}_v$ in the large-$\lambda$ limit
\begin{equation}\label{vamplj}
\begin{split}
\mathcal{A}_v^{j_1,j_2,j_3}&\rightarrow \frac{1}{\pi^7(\lambda\sqrt{1-\gamma^2})^{21}} \Bigg( \frac{e^{\frac{(1+\gamma)}{2}\lambda S_v(\Sigma_1)}}{\sqrt{-D}}+\frac{e^{\frac{(1+\gamma)}{2}\lambda S_v(\Sigma_2)}}{\sqrt{-D^{*}}}\Bigg) \Bigg( \frac{e^{\frac{(1-\gamma)}{2}\lambda S_v(\Sigma_1)}}{\sqrt{-D}}+\frac{e^{\frac{(1-\gamma)}{2}\lambda S_v(\Sigma_2)}}{\sqrt{-D^{*}}}\Bigg)\\
\\
&=\frac{1}{\pi^7(\lambda\sqrt{1-\gamma^2})^{21}} \Bigg( \frac{e^{i\lambda S_{R}}}{-D}+\frac{e^{-i\lambda S_{R}}}{-D^{*}}+2\frac{\cos (\lambda \gamma S_{R})}{\sqrt{D D^{*}}}\Bigg),
\end{split}
\end{equation}
where $S_R$ is the action
\begin{equation}\label{reggea}
S_R(j_1,j_2,j_3)=6 (j_1-j_2) \Big(\frac{\pi}{2}-\theta\Big) + 12 j_3\Big(\frac{\pi}{2}-\arccos(\cos^2\theta)\Big),
\end{equation}
and can be interpreted as the Regge action describing the dynamics of the classical model. Let us observe that it has indeed the form
\begin{equation*}
S_R=\sum_{h}a_h\epsilon_h
\end{equation*}
being $a_h$ the area of the hinge $h$  (i.e., a 2-dimensional face) and $\epsilon_h=\frac{\pi}{2}- \Theta_{h}$ the contribution of the analyzed vertex to the deficit angle at the hinge. The 24 dihedral angles $0<\Theta_{ab}<\pi$ can be computed by performing the scalar product between all the couples  $N_a$,$N_b\in\mathbb{R}^4$ of outward pointing normals to the boundary hexahedra $\varepsilon_a$ and $\varepsilon_b$ (see Appendix). We find six dihedral angles  $\Theta=\theta$ associated to hexahedra which meet along $j_1$ faces, six dihedral angles  $\Theta'=\pi-\theta$ associated to hexahedra meeting along $j_2$ faces and twelve dihedral angles $\Theta''=\arccos(\cos^2\theta)$ corresponding to boundary frusta meeting along  $j_3$ faces.

Let us also notice that both the determinant function \eqref{determH} and the Regge action \eqref{reggea} are invariant under exchange $j_1 \leftrightarrow j_2$. In the light of the physical interpretation which we propose in the next section, a consequence of this symmetry is that the full transition amplitude does not distinguish between space expansions or contractions at the same rate.

Finally, we can absorb the expressions \eqref{fam},\eqref{eamplf} and \eqref{eamplc} of $\mathcal{A}_{f}$ and $\mathcal{A}_{e}$ in the vertex amplitude \eqref{vamplj} in order to write the generating functional \eqref{generatingfunctional} in terms of a dressed vertex amplitude $\hat{\mathcal{A}}_v$.
Since every edge $e$ is bounded by two vertices, we split the contribution of the corresponding edge amplitude by assigning to each vertex sitting at the extremes of $e$ the square root of $\mathcal{A}_{e}$. In the same fashion, since a face is shared by four vertices (corresponding to the fact that, in four dimensions, four 3d hexahedra meet in a 2d trapezoid) we multiply each vertex amplitude with the fourth root of $\mathcal{A}_{f}$. Summarizing, for a generic vertex $v$ we have
\begin{equation}\label{drevampl}
\hat{\mathcal{A}}_{v}\equiv\prod_{f\supset v}\mathcal{A}_{f}^{1/4} \ \prod_{e\supset v}\mathcal{A}_{e}^{1/2} \ \mathcal{A}_{v},
\end{equation}
and the generating functional takes the compact form
\begin{equation}\label{genfin}
Z_{\Gamma}= \sum_{j_{f} , \iota_{e}} \prod_{v} \hat{\mathcal{A}}_{v}.
\end{equation}
\\

This concludes the semiclassical analysis of the EPRL-FK spin foam model in the reduced state sum approximation.
Starting from the above asymptotic formula for the generating functional, one can perform a study of the renormalization properties of the model as well as analyze the restoration of the diffeomorphisms invariance by gauging the parameters of the theory. A preliminary numerical analysis shows that, in the limit in which the spin variables are fixed to reproduce a hypercuboidal lattice, our results are consistent with the one obtained in \cite{Bahr:2015gxa}.
We leave the investigation of these topics for a future research.
%

In the second part of this paper we are going to complete the analysis of the classical properties of the model by focusing on the (Regge-type) action obtained in \eqref{reggea}. We will see how the restricted set of geometrical configurations considered carries enough information to reproduce the standard cosmological dynamics of a flat FLRW universe in the limit of fine discretization of the lattice as well as in the small deficit angles limit.

\section{Modelling Cosmology} \label{modell}

The action \eqref{reggea} encodes the classical properties of the system. It is the generalization of the Regge action to the case of hyperfrusta, instead of triangulations, where the areas (instead of edge lengths) are the free variables. Nonetheless, we will refer to \eqref{reggea} as ``Regge action'' in what follows, and show that, in the limit of large lattices, classical cosmology is obtained. To this end, we investigate the dynamics of the spin variables described by the equations of motion, which we are going to derive in the next section.
In this paper we consider a spacetime manifold $\mathcal{M}\sim T^3 \times [0,1]$ given by the product of the 3-torus and a closed interval. In particular we define homogeneous and locally isotropic states on $T^3$ and let them evolve. Such states are represented by a Daisy graph (see figure \ref{fig:couple} on the left) in which the node is dual to a cube and all the links are labeled by the same spin value. A similar construction has been studied in the context of spin foam cosmology where the transition amplitude between holomorphic coherent states  are calculated \cite{Rennert:2013pfa}.
The dressed vertex amplitude defined in the previous section can be interpreted as the transition probability between two space-like hypersurfaces $\Sigma_i$ and $\Sigma_f$ at different time steps $t_i$ and $t_f$ as it is shown in figure \ref{fig:couple} on the right. In particular, we regard the two cubes at the boundary of a hyperfrustum as isotropic and homogeneous space-like hypersurfaces. The evolution occurs in the bulk region bounded by the six boundary frusta, which in our setup are time-like hypersurfaces.
\begin{figure}[h]
	\centering
	\qquad \qquad
	\includegraphics[scale=0.6]{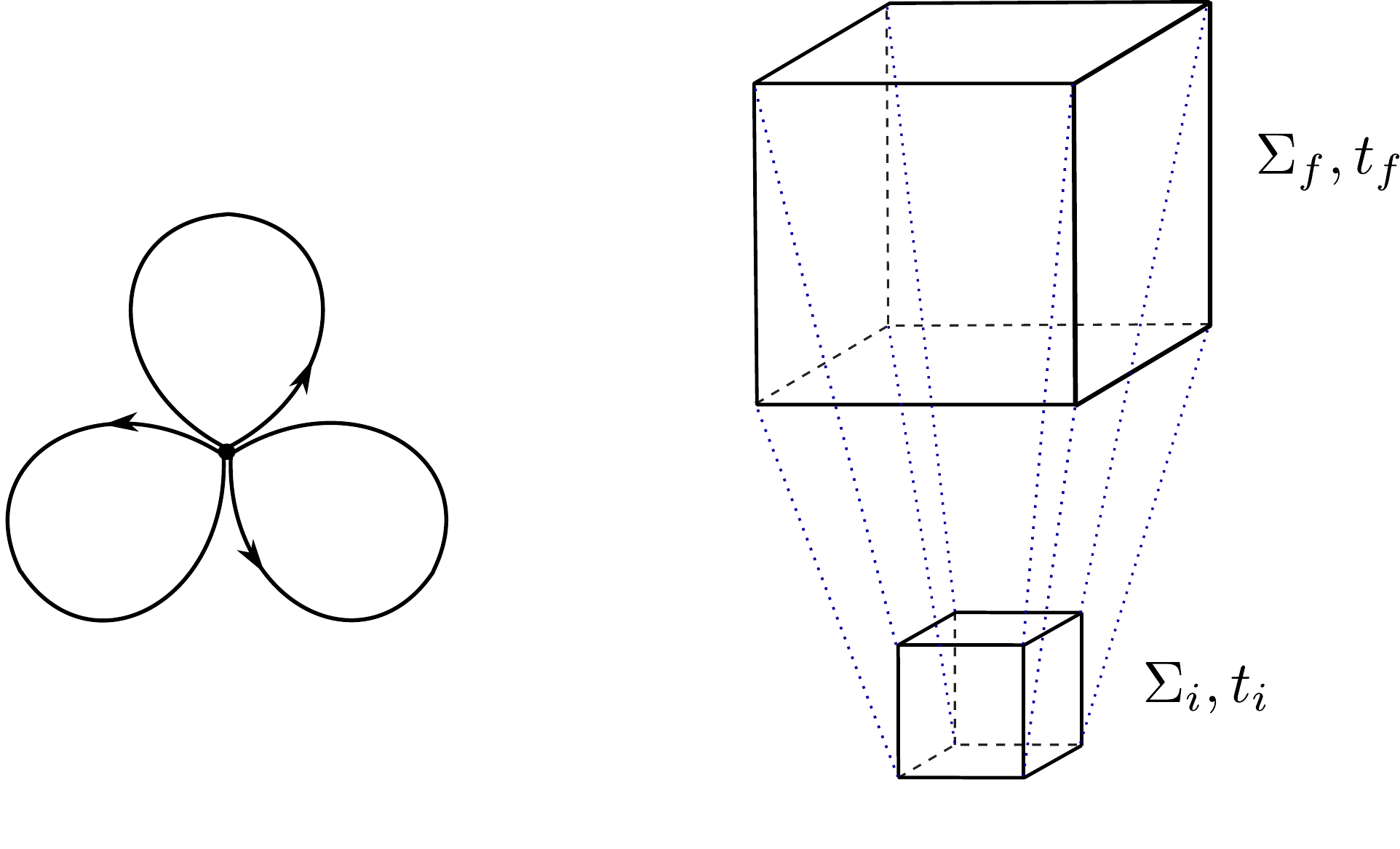}
	\caption{	\small The figure on the left shows a Daisy graph which corresponds to the spin network graph associated to a boundary cube. On the right we represent a hyperfrustum as the time evolution of its space-like boundary cubes $\Sigma_i$ and $\Sigma_f$.
		}
	\label{fig:couple}
\end{figure}
The characteristic size of space at a fixed time is then encoded by the spin values associated to the cube faces. The peculiar choice of reducing the state sum to hyperfrusta makes possible the variation in size of the boundary cubes at  successive time steps. Thus, from an intuitive perspective the model allows a basic concept of expansion and contraction of a flat space.

In order to describe the classical dynamics of the space slices let us consider the chain in figure \ref{fig:line} obtained by gluing together a series of hyperfrusta  $\mathrm{F}_n$ and representing the time evolution of their boundary cubes $\mathrm{c}_n$ having areas $j_n$. At each step the evolution occurs in the bulk region bounded by the six boundary frusta $\mathrm{f}_n$ with bottom faces $j_n$, top faces $j_{n+1}$ and side faces of area $k_n$.
\begin{figure}[h]
	\centering
	\includegraphics[scale=0.66]{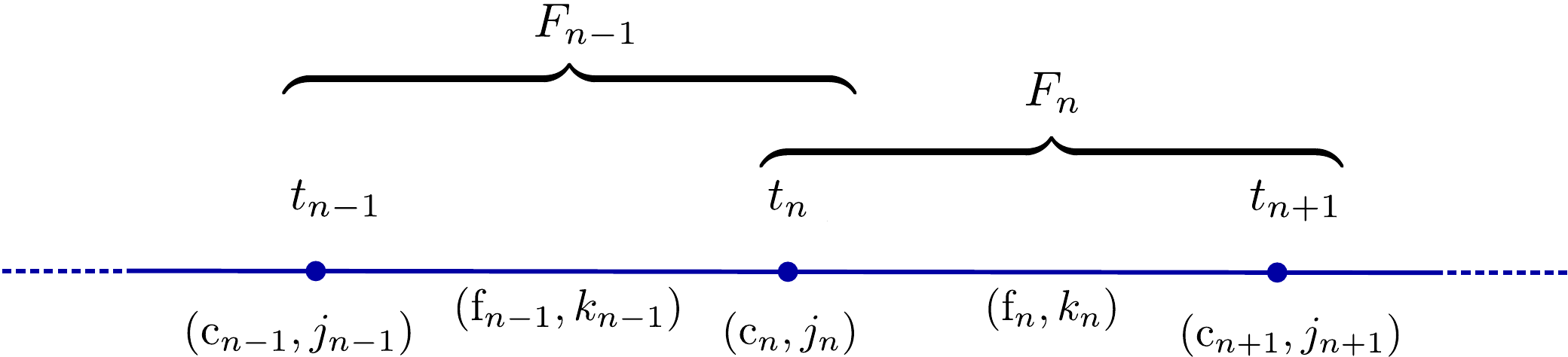}
	\caption{	\small The figure shows a chain obtained by gluing together many hyperfrusta. In particular, the $n$-th node in the chain represents the `past' cube $\mathrm{c}_n$ in the boundary of the hyperfrustum $F_n$. The $(n+1)$-th cube $\mathrm{c}_{n+1}$ is the `future' cube in the boundary of $F_n$. The line connecting these two cubes is associated to the remaining six boundary frusta $\mathrm{f}_n$.}
	\label{fig:line}
\end{figure}
Let us observe that such construction resembles a so-called CW skeleton (Collins-Williams), which is a discrete structure specifically designed to approximate a FLRW universe in the context of Regge calculus \cite{PhysRevD.7.965,Brewin:1987} \footnote{A similar construction is investigated in \cite{Lewis:1982zz} and \cite{Gentle:2012tc} to model the flat FLRW and the Kasner solutions of general relativity.}.
The Cauchy surfaces of a CW skeleton are discretized by regular polytopes (in our case cubes) and, as in the FLRW approximation, they are identical to each other apart from an overall scaling factor. This analogy allows us to interpret the spin $j_n$ associated to the $n$-th cube as a discrete surrogate of the scale factor at a fixed time. Therefore, we define the {\itshape scale factor} at the $n$-th step as
\begin{equation}\label{scalef}
a_n \equiv\sqrt{j_n}.
\end{equation}
Let us also define the time step of the evolution between the cubes $\mathrm{c}_n$ and $\mathrm{c}_{n+1}$ to be the distance between their centers or, equivalently, the height $H_n$ of $F_n$ i.e.,
\begin{equation*}
t_{n+1}-t_n\equiv  H_n.
\end{equation*}
Let $\theta_n$ be the dihedral angle between $\mathrm{c}_n$ and $\mathrm{f}_n$ and let $h_n$ be the height of $\mathrm{f}_n$. From the results of the last section and using arguments of classical geometry one can show that their values in terms of the spins are
\begin{equation}\label{varia}
\begin{split}
\theta_n \ &= \ \arccos(\cot\phi_n),\\
h_n \ &= \ \frac{2k_n}{\sqrt{j_{n+1}}+\sqrt{j_n}}\sin \phi_n,
\end{split}
\end{equation}
being $\phi_n$ the slope angle of the frustum $\mathrm{f}_n$ such that (in analogy with \eqref{anglespin})
\begin{equation*}
\cos \phi_n=\frac{j_{n+1}- j_n}{4k_n}.
\end{equation*}
In terms of these variables we find the expression for the $n$-th time step
\begin{equation}\label{heighthyperfr}
H_n=h_n\sin \theta_n = \frac{2k_n}{\sqrt{j_{n+1}}+\sqrt{j_n}}  \ \sqrt{1-\frac{(j_{n+1}-j_n)^2}{8k_n^2}}.
\end{equation}

Before proceeding to the explicit computation of the equations of motion, let us find out how the vacuum Friedmann equations look like in the reduced model under study by performing a qualitative analysis.
From the above definitions we can compute the discrete time derivative of the scale factor as a function of the spin variables
\begin{equation*}
\dot{a}_n=\frac{a_{n+1}-a_n}{t_{n+1} - t_n}= \frac{2}{\sqrt{\tan^2\phi_n-1}}.
\end{equation*}
Using the first equation in \eqref{varia} it is easy to check that, in terms of the dihedral angle $\theta_n$ between $\mathrm{c}_n$ and $\mathrm{f}_n$, the expression above reads
\begin{equation}\label{adot}
\dot{a}_n= 2 \cot \theta_n.
\end{equation}
The first vacuum Friedman equation $\dot{a}_n=0$ would then tell us that locally the classical evolution happens for $\phi_n=\frac{\pi}{2}$ i.e., on a hypercubic lattice in which all the dihedral angles are $\theta_n=\frac{\pi}{2}$.  Since at each square in the lattice the contribution to the deficit angle is given by four hypercuboids, then the sum of the angles vanishes at all the hinges, which corresponds to flat space.
The second derivative of the scale factor is easily derivable and reads
\begin{equation}\label{addot}
\ddot{a}_n=- \frac{2}{\sin^2\theta_n} \Big(\frac{\theta_n-\theta_{n+1}}{t_{n+1}-t_n}\Big).
\end{equation}
Since $1/2\leq \sin \theta_n \leq 1$ is constrained by  the consistency condition \eqref{allowedspins} and is not vanishing, we deduce that the acceleration of the scale factor vanishes only when the dihedral angle does not vary with the time flow i.e., $\theta_n=\theta_{n+1}$.
Therefore, at the scale defined by the building blocks, the vacuum Friedman equations $\ddot{a}_n=\dot{a}_n=0$ are fulfilled only in the case of a flat reduced universe with vanishing deficit angles at the hinges. Let us note that in general an accelerated expansion (contraction) of the universe would be described by a growth (decrease) of the dihedral angles at successive steps. The next step in our analysis is the explicit derivation of the equations of motion. In fact we want to verify that the expected results are obtained without imposing the Friedmann equations a priori as we just did.

\section{Dynamics Of The Model} \label{dyn}
We are now going to study the classical dynamics of the discrete model by deriving the equations of motion for the action (\ref{reggea}) in three cases: pure gravity, in presence of a cosmological constant and in the case of dust matter coupling.
It is known that, given a generic triangulation, difficulties may arise in the context of Regge calculus when considering two-dimensional areas as independent variables instead of the edge lengths \cite{Barrett:1997tx}.
In particular, the information given by the areas of a four dimensional polyhedron is in general not enough to unambiguously reconstruct its geometry. For example, although a 4-simplex has the same number of edge lengths and faces, one can construct two 4-simplices with the same triangular areas but different edge lengths. The situation gets worse in the case of many four-dimensional blocks glued together. Another ambiguity is in the interpretation of the Regge equations where, for instance, the vanishing of the deficit angles (seen as functions of the areas) does not necessarily imply flatness. Various solutions to these issues have been studied in the literature \cite{Makela:1993ed,Makela:1998wi,Makela:2000ej}, and extensions of the so-called area Regge calculus have been proposed \cite{Dittrich:2008va}.
These concerns, however, are not necessary in the context of our model where the spins are a priori constrained into a rigid symmetric configuration.
In fact, the number of spins required to reconstruct the geometry of a regular hyperfrustum is equal to the number of independent edge lengths.  Further, this result holds for arbitrary numbers of hyperfrusta glued together.
As a consequence, one can freely invert the relationship between length and spin variables without affecting the accuracy of the geometrical description.
Finally, as we will see, the equations of motion derived are equal to the standard Regge calculus ones.
The following analysis is inspired by a collection of works on cosmological models with Regge calculus \cite{PhysRevD.7.965,Lewis:1982zz,Brewin:1987,Liu:2015bwa,Liu:2015gpa,Liu:2015whw,Tsuda:2016ryp,Gentle:2012tc}.
\subsection{Flat vacuum FLRW universe}
Let us refer once again to the chain model in figure \ref{fig:line}.
The full Regge action is given by a sum of terms of the form \eqref{reggea} for each hyperfrustum $F_n$
\begin{equation}
S_R(\{j_n\},\{k_n\})=\sum_{n} S_{R,n}(j_n,j_{n+1},k_n)=\sum_{n} \Bigg( \frac{3}{2}(j_n-j_{n+1}) \delta^{(j)}_{n}+3k_{n}\delta^{(k)}_{n}\Bigg),
\end{equation}
 being the deficit angles
 \begin{equation}
 \begin{split}
 \delta^{(k)}_{n}&=2\pi -4\arccos (\cos^2\theta_n),\\
 \delta^{(j)}_{n}&=2\pi -4\theta_n,
 \end{split}
 \end{equation}
 and
 \begin{equation}\label{thetangle}
 \cos \theta_n=\frac{j_{n+1}-j_n}{\sqrt{16 k_n^2-(j_{n+1}-j_n)^2}}.
 \end{equation}
 Deriving the Regge action with respect to the spins $k_n$ and $j_n$ and setting the result equal to zero gives the equations of motion which solve the classical dynamics of the discrete model \footnote{ Such procedure is regarded as a global variation since the six spins $j_n$ of $c_n$, as well as the twelve spins $k_n$ of the frusta $\mathrm{f}_n$, are first constrained to form a regular hyperfrustum and then they are all derived at once. A local variation would instead consider each spin separately and impose the constraints at the end. For more details see \cite{Brewin:1987}.}.
A direct calculation shows that the contribution of the derivatives of the dihedral angles sum up to zero. Thus, a posteriori, one does not need to derive the deficit angles in the Regge action in order to obtain the equations of motion. This can be regarded as the analogue of the Schl{\"a}fli identity \cite{math/0001176}. The Regge equations of motion for the spins $k_n$ and $j_n$ are then
\begin{equation}\label{reggeq}
\begin{split}
\frac{\partial S_R}{\partial k_n}&=3  \ \delta^{(k)}_{n}=0,\\
\frac{\partial S_R}{\partial j_n}&=\frac{3}{2} \ \Big(\delta^{(j)}_{n}-\delta^{(j)}_{n-1}\Big)=0.
\end{split}
\end{equation}
Let us notice that these equations correspond respectively to the vanishing of \eqref{adot} and \eqref{addot}. Indeed, the first equation of motion implies the vanishing of the dihedral angle $\theta_n$, while the second equation tells us that the dihedral angle remains constant at successive time steps i.e., $\theta_n=\theta_{n+1}$.
Therefore, as it is illustrated in the previous section, the equations of motion \eqref{reggeq} can be interpreted as a discrete version of the vacuum Friedmann equations.

Let $m_n$ be the length of a `strut' of the $n$-th frustum (i.e., the diagonal edge of its trapezoidal faces) and $l_n$ the edge length of the $n$-th cube. One can show that the first equation in \eqref{reggeq} is equivalent to the one obtained by deriving the Regge action w.r.t.~the strut length, apart from an overall non-vanishing factor. Explicitly,
\begin{equation}
\frac{\partial S_R}{\partial m_n}= \frac{\partial k_n}{\partial m_n} \frac{\partial S_R}{\partial k_n}=0.
\end{equation}
It has been noted that such equation can be interpreted as the analogue of the Hamiltonian constraint of the ADM formalism \cite{Brewin:1987}.
In the same way, the equation of motion for the variable $l_n$ is linked to the evolution equation of ADM formalism and it can be written as
\begin{equation}
\frac{\partial S_R}{\partial l_n}=
\frac{\partial j_n}{\partial l_n} \frac{\partial S_R}{\partial j_n}
+\frac{\partial k_n}{\partial l_n} \frac{\partial S_R}{\partial k_n}
+\frac{\partial k_{n-1}}{\partial l_n} \frac{\partial S_R}{\partial k_{n-1}}=0.
\end{equation}
This coincides with the equation of motion for the spin $j_n$ only when it is evaluated on the solution of the equations of motion for the variables $k_n$ and $k_{n-1}$.
We will still refer to $\partial S_R/\partial k_n=0$  as the {\itshape Hamiltonian constraint} and to $\partial S_R/\partial j_n=0$  as the {\itshape evolution equation}. Such observations will be valid also in the next subsections where we study the Friedmann universe in presence of a cosmological constant and coupled to dust particles.

In order to remove any doubt about the connection between the Regge equations of motion \eqref{reggeq} and the vacuum Friedmann equations, let us pass to the continuum {\itshape time} limit.
From the time step formula \eqref{heighthyperfr} we get
\begin{equation}\label{kappan}
k_n^2= \frac{(\sqrt{j_{n+1}}+\sqrt{j_n})^2}{4}H_n^2 +  \frac{(j_{n+1}-j_n)^2}{8}.
\end{equation}
Substituting this expression into the dihedral angle \eqref{thetangle}, one can write the Regge equations \eqref{reggeq} in terms of the spins $j_n$'s and the time steps $H_n$'s.
Let us now perform the following replacement in the equations of motion
\begin{equation}
\begin{split}
H_n, H_{n-1} &\rightarrow \mathrm{d}t,\\
j_n &\rightarrow j(t),\\
j_{n+1} &\rightarrow j(t) + j' \mathrm{d}t + \frac{1}{2}j'' \mathrm{d}t^2 + \mathcal{O}(\mathrm{d}t^3),\\
j_{n-1} &\rightarrow j(t) - j' \mathrm{d}t + \frac{1}{2}j'' \mathrm{d}t^2+\mathcal{O}(\mathrm{d}t^3),
\end{split}
\end{equation}
and find the continuum time limit by sending $\mathrm{d}t\rightarrow0$. Note that we have imposed that the time step $H_n$ is constant in this limit $\forall n$. This corresponds to a gauge fixing choice and it is justified by the fact that the equations of motion \eqref{reggeq} do not impose constraints on the allowed values  of $k_n$ and $H_n$.
At the leading order in $\mathrm{d}t$ the Regge equations read
\begin{equation}\label{eomm}
\begin{split}
3\Bigg(2\pi -4\arccos\frac{j'^2}{16 j+j'^2}\Bigg)&=0,\\
12 \ \frac{1}{\sqrt{j}} \frac{2j j''-j'^2}{16j + j'^2} &=0.
\end{split}
\end{equation}
Deriving the Hamiltonian constraint (first equation) one can easily check that it is a first integral of the evolution equation (second equation).
Let us note that we are still working in Euclidean signature. To argue a solution which is comparable to the standard Friedmann cosmology we need to perform a Wick rotation $t \rightarrow i t$. This results in the replacements $j''\rightarrow-j''$ and $j'^2\rightarrow-j'^2$. One can check that the vacuum solutions remain unchanged. However, this step will be fundamental when investigating the coupling to cosmological constant and to dust particles.
The solutions of the Hamiltonian constraint and the evolution equation are readily derived
\begin{equation}
j'=0, \qquad \qquad j''=\frac{j'^2}{2j}.
\end{equation}
In the interpretation given in the previous section in which the scale factor is $a=\sqrt{j}$, the Regge equations correspond to
\begin{equation}
\begin{split}
\frac{a'^2}{a^2}&=0,\\
\frac{a''}{a}&=0,
\end{split}
\end{equation}
which are the standard vacuum Friedmann equations for a flat universe.

\subsection{Flat $\Lambda$-FLRW universe}
The action in presence of a cosmological constant term $\Lambda>0$ is
\begin{equation}\label{reggg}
	S_R(\{j_n\},\{k_n\},\Lambda)= \sum_{n}\Bigg( \frac{3}{2}(j_n-j_{n+1}) \delta^{(j)}_{n}+3k_{n}\delta^{(k)}_{n}-\Lambda V_n\Bigg),
\end{equation}
being $V_n$ the four dimensional volume of the $n$-th hyperfrustum. We can express it in terms of the spins as (see Appendix)
\begin{equation}\label{volume}
V_n=  \frac{k_n  (j_n + j_{n+1})}{2} \ \sqrt{1-\frac{(j_{n+1}-j_n)^2}{8k_n^2}}.
\end{equation}
The new Regge equations are
\begin{equation}
\begin{split}
\frac{\partial S_R}{\partial k_n}&=3  \ \delta^{(k)}_{n}-\Lambda\frac{\partial V_n}{\partial k_n}=0,\\
\frac{\partial S_R}{\partial j_n}&=\frac{3}{2} \ \Big(\delta^{(j)}_{n}-\delta^{(j)}_{n-1}\Big)-\Lambda\Bigg( \frac{\partial V_n}{\partial j_n}+\frac{\partial V_{n-1}}{\partial j_n}\Bigg)=0.
\end{split}
\end{equation}
Performing the continuum time limit as we did in the vacuum case, one can find the Hamiltonian constraint and the evolution equation for a flat $\Lambda$-FLRW universe. After a Wick rotation $t\rightarrow i t$, $j''\rightarrow-j''$,  $j'^2\rightarrow-j'^2$ they read
\begin{equation}\label{hamevo}
\begin{split}
2\pi -4\arccos\frac{j'^2}{j'^2-16 j} \ &= \ \frac{\Lambda}{3}j \sqrt{1-\frac{j'^2}{8j}},\\
\frac{2j j''-j'^2}{j'^2-16 j} \ &= \ \frac{\Lambda}{12}j \Bigg(1-\frac{j'^2}{16j}-\frac{j''}{8} \Bigg).
\end{split}
\end{equation}
As in the vacuum case, the Hamiltonian constraint (first equation) is the first integral of the evolution equation (second equation). Thus, we can use it to study the evolution of the model.
Notice that the Hamiltonian constraint is only defined for
\begin{equation}\label{co}
\frac{j'^2}{8j}\leq 1,
\end{equation}
which imposes a condition on the maximal rate of expansion of the space surfaces.
Let us define the Wick-rotated dihedral angle associated to the time-like hinges
\begin{equation}\label{wicktheta}
\Theta_W\equiv \arccos\frac{j'^2}{j'^2-16 j}.
\end{equation}
When evaluated in the range \eqref{co} this is a function with real values
\begin{equation}\label{a}
 \frac{\pi}{2} \leq \Theta_W\leq \pi, \qquad \quad -1\leq \cos \Theta_W\leq 0.
\end{equation}
From \eqref{wicktheta} we find
\begin{equation}\label{jprime}
j'^2=-16 j \frac{\cos \Theta_W}{1-\cos \Theta_W}.
\end{equation}
 Using the above definitions the Hamiltonian constraint becomes
\begin{equation}\label{jpara}
j^2=\frac{9}{\Lambda^2} \ \frac{1-\cos \Theta_W}{1+\cos \Theta_W} \ (2\pi -4\Theta_W)^2.
\end{equation}

%
%
%
Expressing the volume of the universe as $U=j^{3/2}$ we can find the equation describing its time evolution
\begin{equation}
 \frac{\mathrm{d}U}{\mathrm{d}t}= \frac{3}{2}j^{\frac{1}{2}}j'= 6 j \sqrt{\frac{-\cos \Theta_W}{1-\cos \Theta_W}}\\
=-\frac{18}{\Lambda} \ \sqrt{\frac{-\cos \Theta_W}{1+\cos \Theta_W}} \ (2\pi -4\Theta_W).
\end{equation}
where we have used the equations \eqref{wicktheta} and \eqref{jprime}. Let us notice that also the square root of equation \eqref{jpara} is involved in the above derivation. Since it can assume both positive and negative values, one must carefully select the signs according to the angle range \eqref{a} in order to get a positive value of $j$.

The volume and its time variation form a set of parametric equations which can be solved using numerical methods.

\begin{figure}[h]
\qquad 	\includegraphics[scale=0.56]{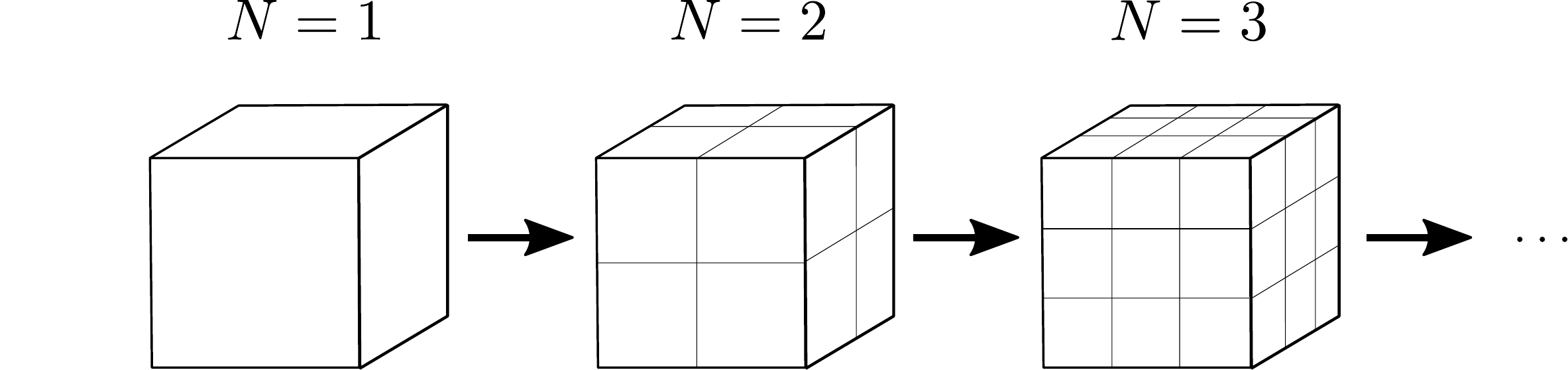}
	\caption{\small The figure shows some coarse graining steps of a 3-Torus.}
	\label{fig:coarse}
\end{figure}
Note that the use of a rigid hyperfrustum is not well suited to capture the degrees of freedom of a constantly curved spacetime such as in the case of a Friedmann universe in presence of a cosmological constant. Thus, in order to get a better approximation of the Friedmann dynamics one needs to refine the lattice discretization by describing the evolution of a larger number of cubes tessellating each Cauchy surface \footnote{Actually, one can also consider the use of constantly curved building blocks to discretize spacetime as in \cite{Bahr:2009qc,Bahr:2009qd}} as in figure \eqref{fig:coarse}.
In the case we want to describe the evolution of $N^3$ identical cubes, the Hamiltonian constraint does not vary since the number of cubes factorizes in the action \eqref{reggg} and the continuum time limit procedure is not affected by the coarse graining. What changes is instead (modulo rescaling) the volume of the universe
\begin{equation}
U\rightarrow N^3 U.
\end{equation}
In figure \eqref{fig:plot} we plot the time derivative of the volume (for some positive value of $\Lambda$) against the volume of the universe itself for different numbers of cubes tessellating a Cauchy surface. The results are compared to the analytic ones obtained from the Friedmann equations of a flat universe with cosmological constant i.e.,
\begin{equation}
U_{\mathrm{analytic}}=a^3=e^{\sqrt{3\Lambda}t}, \qquad \qquad
\frac{\mathrm{d}U_{\mathrm{analytic}}}{\mathrm{d}t}=\sqrt{3\Lambda} \ e^{\sqrt{3\Lambda}t}.
\end{equation}

\begin{figure}[h]
	\ \ \  \
	\begin{overpic}[scale=0.63,unit=1mm]{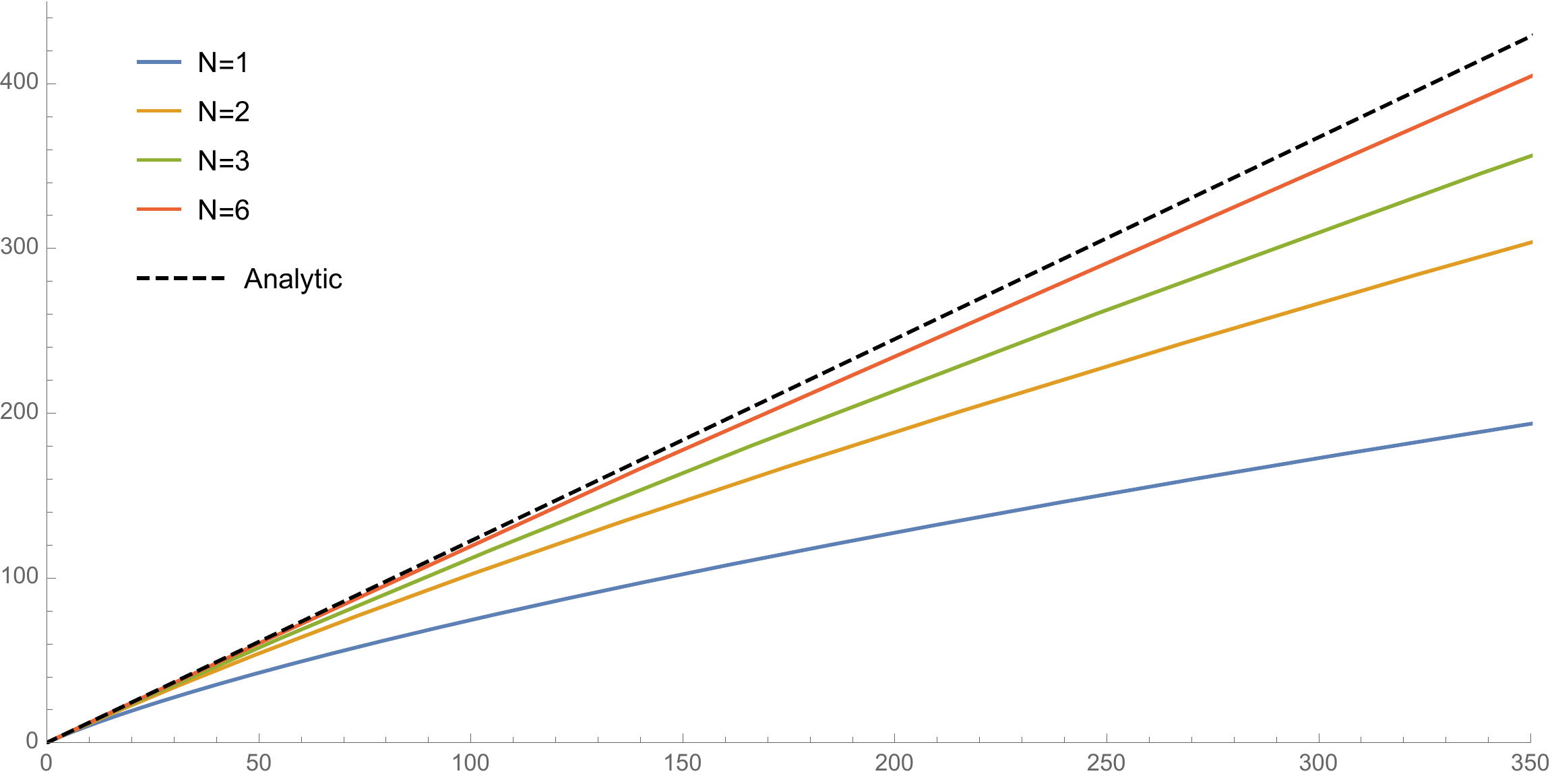}
		\put(28.5,44){{\parbox{0.4\linewidth}{%
					\begin{equation*}
						\Lambda = 0.5
					\end{equation*}}}}
					\put(-23,26){{\parbox{0.4\linewidth}{%
								\begin{equation*}
									\frac{\mathrm{d}U}{\mathrm{d}t}
								\end{equation*}}}}
								\put(28.5,-2){{\parbox{0.4\linewidth}{%
											\begin{equation*}
												U
											\end{equation*}}}}
										\end{overpic}
	\\
	\caption{Flat universe with cosmological constant $\Lambda = 0.5$, as approximated by the hyperfrustal evolution with $N^3$ cubes.}
	\label{fig:plot}
\end{figure}

In many models which make use of the CW formalism the Cauchy surfaces analyzed are 3-spheres triangulated by using {\itshape regular} tetrahedra. Therefore the universe examined is a closed one. However a 3-sphere can be triangulated by using only 5, 16 or 600 regular tetrahedra (see for example \cite{Coxeter:1948}), thus there exists a geometric constraint which prevents from approaching the analytic limit at will.
The advantage of the model studied in this article is that a flat $3$-torus can be tessellated by an arbitrarily high number of cubes and there is no theoretical limit to the refinement steps that one can take to show the convergence to the analytic results.

Another way to solve the Hamiltonian constraint is by studying the limit in which the deficit angle at the hinges is small, corresponding to a slow (measured in Planck times) expansion or contraction of the universe. In fact, only in this regime the discrete lattice of Regge calculus approximates the continuous smooth manifold of general relativity \cite{Regge:1961px}. In our case such limit is made explicit by the requirement
\begin{equation}\label{k1}
\Theta_W =\frac{\pi}{2}+\eta, \qquad |\eta| \ll 1.
\end{equation}
Intuitively this condition indicates that the boundary frusta $\mathrm{f}_n$ in figure \ref{fig:line} present a small deviation from a cuboidal geometry.
Substituting the above expression into equation \eqref{jprime} and taking the limit $\eta \rightarrow 0$ we find at the leading order of $\eta$ and $j'^2$, that
\begin{equation}\label{eps}
\eta=\frac{j'^2}{16 j}.
\end{equation}

Let us come back to the Hamiltonian constraint \eqref{jpara} and substitute the angle \eqref{eps}. We get
\begin{equation}
j^2=\frac{9}{\Lambda^2}  \ (-4\eta)^2= \frac{9}{\Lambda^2} \ 16 \Bigg(\frac{j'^2}{16 j}\Bigg)^2
\end{equation}
Finally, from the definition of the scale factor $a=\sqrt{j}$ we obtain the first Friedmann equation for a flat $\Lambda$-FLRW universe
\begin{equation}
a'^2=\frac{\Lambda}{3}a^2.
\end{equation}
The second Friedmann equation is simply given by the time derivative of the first and reads
\begin{equation}
a''=\frac{\Lambda}{3}a.
\end{equation}
Let us note that this is consistent with the fact that the evolution equation is the derivative of the Hamiltonian constraint in \eqref{hamevo}. In fact, one can check that the second Friedmann equation can also be derived from the evolution equation using the same arguments just presented.


\subsection{Flat FLRW universe with dust}
Let us place a test particle of mass $M$ at the center of each cube $c_n$ in the chain \eqref{fig:line}. Classically, the motion of a point particle in a gravitational field is found by applying the variational principle to the following action
\begin{equation}
S_M=-M\int \mathrm{d}s,
\end{equation}
being $\mathrm{d}s$ the line element. We define the discrete analogue of the line element as the length $s_n$ of the trajectory joining the centers of the cubes $c_n$ and $c_{n+1}$. In fact the choice of placing the test particle at the center of the cubes guarantees that it is comoving and travels along geodesics \cite{Liu:2015bwa}. Thus in our case the discrete line element is given by the time step \eqref{heighthyperfr} i.e., $s_n=H_n$ (remember that we are working in Euclidean signature). More general settings have been studied on a simplicial discretization of a closed universe. For example, in \cite{Liu:2015bwa} it has been shown that the Hamiltonian constraint depends on the particle position inside the tetrahedra. In the following analysis we are going to describe the evolution of a single test particle in spacetime.

In order to describe a universe in which more than one dust particle is present, one can refine the lattice as in \ref{fig:coarse} and distribute $N^3$ particles, each of mass $M/N^3$, over the initial cubes, such that one particle sits at the center of each cube.
The full action becomes
\begin{equation}\label{actM}
	S_R(\{j_n\},\{k_n\},M)= \frac{N^3}{8 \pi} \sum_{n}\Bigg( \frac{3}{2}(j_n-j_{n+1}) \delta^{(j)}_{n}+3k_{n}\delta^{(k)}_{n}\Bigg)- M \sum_{n} H_n,
\end{equation}
where we have rehabilitated the factor $1/8\pi$ in front of the Regge action\footnote{In the previous cases the factor $1/8\pi$ does not contribute to the dynamics since it factorizes in the action.} and we are working in Plank units $c=G=1$.
The new Regge equations are
\begin{equation}
	\begin{split}
		\frac{\partial S_R}{\partial k_n}&=\frac{3N^3}{8 \pi}  \ \delta^{(k)}_{n}-M\frac{\partial H_n}{\partial k_n}=0,\\
		\frac{\partial S_R}{\partial j_n}&=\frac{3N^3}{16 \pi} \ \Big(\delta^{(j)}_{n}-\delta^{(j)}_{n-1}\Big)-M\Bigg( \frac{\partial H_n}{\partial j_n}+\frac{\partial H_{n-1}}{\partial j_n}\Bigg)=0.
	\end{split}
\end{equation}
Performing the continuum time limit and the Wick rotation we get the Hamiltonian constraint and the evolution equation
\begin{equation}
	\begin{split}
		2\pi -4\arccos\frac{j'^2}{j'^2-16 j} \ &= \ \frac{8\pi M}{3N^3}\frac{1}{\sqrt{j}} \sqrt{1-\frac{j'^2}{8j}},\\
		\frac{2j j''-j'^2}{j'^2-16 j} \ &= \ -\frac{\pi M}{3N^3}\frac{1}{\sqrt{j}} \Bigg(1-\frac{j'^2}{4j}+\frac{j''}{4} \Bigg).
	\end{split}
\end{equation}
Once again, it is easy to check that the second equation is the time derivative of the first.
Substituting equation \eqref{jprime} in the Hamiltonian constraint and applying the Wick rotation one gets
\begin{equation}\label{gei}
	j=\Bigg(\frac{8\pi M}{3N^3}\Bigg)^2 \ \frac{1+\cos \Theta_W}{1-\cos \Theta_W} \ \frac{1}{(2\pi -4 \Theta_W)^2},
\end{equation}
where the Wick-rotated angle $\Theta_W$ is given in \eqref{wicktheta}.
From the above equation we can write the set of parametric equations describing the volume of the universe and its time variation
\begin{equation}\label{u}
 U=N^3 j^\frac{3}{2}, \qquad \qquad \frac{\mathrm{d}U}{\mathrm{d}t}=6N^3  j \sqrt{\frac{-\cos \Theta_W}{1-\cos \Theta_W}}.
\end{equation}

 The Friedmann equations describing the evolution of the scale factor $a(t)$ in a flat space and in presence of dust are
\begin{equation*}
\begin{split}
&\frac{\dot{a}^2}{a^2}=\frac{8\pi}{3}\rho, \\
&\frac{\ddot{a}}{a}=-\frac{4\pi}{3} \rho,
 \end{split}
 \end{equation*}
 being $\rho=M/a^{\frac{3}{2}}$ the density of the universe.
 Using the same arguments that we applyied in the cosmological constant case, one can check that the above equations can in fact be obtained as the small deficit angle limit of the Hamiltonian constraint and the evolution equation.
 Their solution is
 \begin{equation}\label{solfri}
 a(t)=(6\pi M)^{\frac{1}{3}} t^{\frac{2}{3}},
 \end{equation}
thus the analytic expression for the volume of the universe and its time variation are
\begin{equation*}
U_{\mathrm{analytic}}=a^3=6\pi M t^2, \qquad \qquad
\frac{\mathrm{d}U_{\mathrm{analytic}}}{\mathrm{d}t}=12 \pi M t.
\end{equation*}
 For some value of the mass $M$ we can plot the numerical result \eqref{gei},\eqref{u} to find that the model converges quite rapidly to the analytic curve (see figure \ref{fig:plot1}).
\begin{figure}[h]
	\ \ \ \
	\begin{overpic}[scale=0.63,unit=1mm]{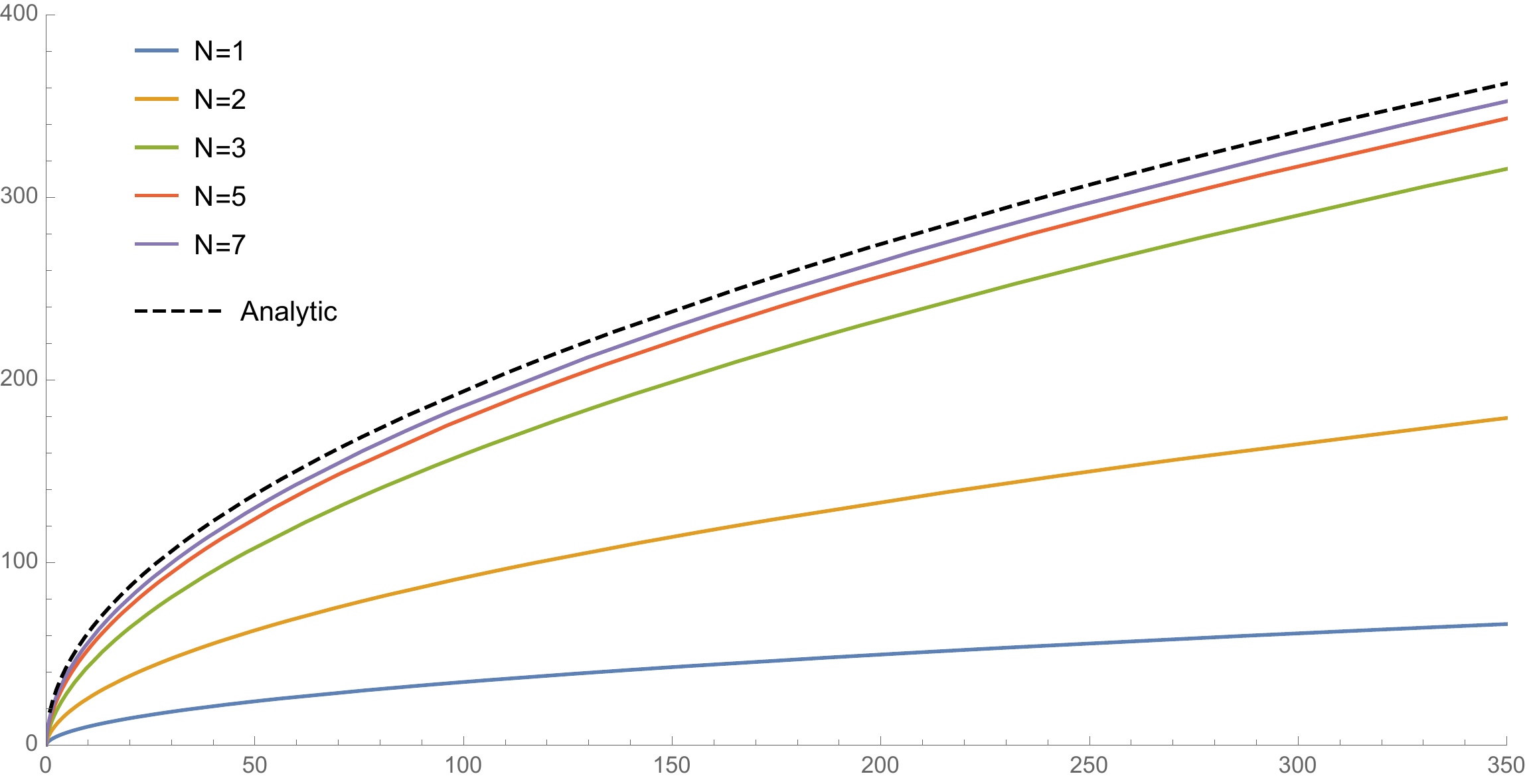}
		\put(28,45){\colorbox{white}{\parbox{0.4\linewidth}{%
					\begin{equation*}
						M = 5
					\end{equation*}}}}
					\put(-24,28){{\parbox{0.4\linewidth}{%
								\begin{equation*}
									\frac{\mathrm{d}U}{\mathrm{d}t}
								\end{equation*}}}}
								\put(28,-2){{\parbox{0.4\linewidth}{%
											\begin{equation*}
												U
											\end{equation*}}}}
										\end{overpic}
	\\
	\caption{Evolution of flat universe filled with dust of mass $M$, with space approximated by $N^3$ cubes.}
	\label{fig:plot1}
\end{figure}

%
%

\section{Summary And Conclusions}
In the first part of this article we investigate the semiclassical structure of the Euclidean EPRL-FK model of quantum gravity.
We work on a hypercubic lattice in which all the vertices are dual to a 4d truncated pyramid with cubic bases (hyperfrustum). Furthermore we restrict the state sum by considering only coherent intertwiners which in the large-spin limit reproduce the geometry of a 3d pyramidal frustum.
The reduced state sum allows us to compute explicitly the asymptotic formula of the vertex amplitude. We show that the final expression contains the correct Regge action describing the classical properties of our model.
Starting from the expression \eqref{genfin} one can perform a numerical analysis of the semiclassical features of the model. This result opens a path to study the renormalization of a reduced model of quantum gravity. Further developments are possible in studying the diffeomorphisms symmetry which is usually broken by the discretization of the spacetime manifold. In fact, knowing the full analytic expression of the partition function and gauging the parameters in the theory one can look for configurations in which this symmetry is restored. Such perspectives can potentially shed a new light on a sector of the EPRL model which is still vastly unexplored. This research line has been originally paved by a series of works on hypercuboidal geometry and non-trivial results have been found in the case of flat spacetime \cite{Bahr:2015gxa,Bahr:2016hwc,Bahr:2017klw}. This article is the first step further in this direction. Future analysis may take into account more general settings which enlarge the number of degrees of freedom in the path integral broaden the set of allowed geometrical configurations. In particular, in the purely flat case only $\alpha$, as set in the face amplitude, is a running coupling constant, while the inclusion of hyperfrusta also offers Newton's constant and/or the Barbero-Immirzi parameter as nontrivial coupling. This will make the renormalization computation much more general. We postpone the study of these aspects for future research.

In the second part of the article, taking inspiration from a series of works on cosmological modelling with Regge calculus \cite{PhysRevD.7.965,Lewis:1982zz,Brewin:1987,Liu:2015bwa,Liu:2015gpa,Liu:2015whw,Tsuda:2016ryp,Gentle:2012tc}, we have completed the study of our model by focusing on its classical description.
We have first shown that the discretization of spacetime in terms of hyperfrusta have a clear classical interpretation. In fact, a hyperfrustum can be pictured as the time evolution of its boundary cubes, each of them tessellating a flat Cauchy surface. The regular geometry of the cubes and their even distribution on the lattice reproduce an isotropic and homogeneous space. Moreover, the change in size of the cubes in the boundary of a hyperfrustum mimic an expansion of the universe. These facts enable us to compare the dynamics of our model to the FLRW one. We do it in three different cases: In the vacuum, in presence of a cosmological constant and by coupling dust particles to the lattice. The simplicity of our model allows us to consider the spins as the main variables instead of the edge lengths which are usually adopted in Regge calculus. Notably, the results do not change and an analogue of the Schl{\"a}fli identity is proved to be satisfied.  Indeed, with a numerical analysis of the Regge equations (for the spins), we show that in the continuum time limit the evolution of the model universe resembles the one predicted by the standard Friedmann dynamics in the case of fine discretization of the manifold. Furthermore, for small deficit angles this resemblance becomes exact and we find the Friedmann equations as the limit of the Regge equations.
%

A crucial open question is, of course, in what way this model can be used to perform actual quantum cosmological computations. Apart from the signature issues, the first quantum correction of this model comes from the Hessian matrix. This matrix is, in general, complex, such that its phase would give quantum corrections to the Regge action, while its modulus provides the path integral measure. It would be quite interesting to see whether these corrections have a classical limit which can be interpreted as higher order terms in the Einstein-Hilbert action. To probe the deep quantum regime in order to derive e.g.~statements about singularity avoidance, however, one would have to depart from the large-$j$ asymptotics, and consider the full amplitude in the regime of small spins. We will return to these issues in future publications.

\section*{Acknowledgements}

The authors would like to thank Sebastian Steinhaus for valuable discussions. This work was funded by the project BA 4966/1-1 of the German Research Foundation (DFG).

\section{Appendix}
Here we derive some geometric properties of the hyperfrustum. Although all the formulas that we are going to derive can be found by assigning a set of four dimensional coordinates to the elements in the boundary of the hyperfrustum, we will propose different solutions which do not require this labeling.
\subsection{Dihedral angles}
The dihedral angles $\Theta_{ab}$ between the couples of hexahedra in the boundary of a hyperfrustum (as depicted in figure \ref{fig:trabound}) can be found from the critical points in table \ref{tab1} using the following formula (see \cite{Barrett:2009gg})
\begin{equation}
\cos \Theta_{ab}= N_a \cdot N_b= \frac{\chi_{ab}}{2}\mathrm{tr}\Big[ g_a(\Sigma_1) g_a(\Sigma_2)^{-1} g_b(\Sigma_2) g_b(\Sigma_1)^{-1} \Big],
\end{equation}
being $N_a$ and $N_b$ the four dimensional outward-pointing normals to the hexahedra $a$ and $b$ and
\begin{equation*}
\chi_{ab}=
\left\{
\begin{array}{rl}
1 & \mbox{if } a,b\in[0,3] \\
1 & \mbox{if } a,b\in[4,7] \\
-1 & \mbox{if } a\in[0,3] \wedge b\in[4,7]\\
-1 & \mbox{if } a\in[4,7] \wedge b\in[0,3]
\end{array}
\right.
\end{equation*}
Notice that this prefactor is necessary since imposing the condition \eqref{impos} we have chosen outward-pointing normals to describe the hexahedra $a=0,1,2,3$ and inward-pointing normals to describe the hexahedra $a=4,5,6,7$.
We find six dihedral angles  $\Theta=\theta$ associated to hexahedra which meet along the faces of the cube $0$, six dihedral angles  $\Theta'=\pi-\theta$ associated to hexahedra meeting along the faces of the cube $7$ and twelve dihedral angles $\Theta''=\arccos(\cos^2\theta)$ corresponding to boundary frusta meeting along  their side faces.
These angles are the four dimensional analogue of the one represented in figure \ref{fig:geometry} on the left.

\begin{figure}[h]
	\quad \
	\includegraphics[scale=0.95]{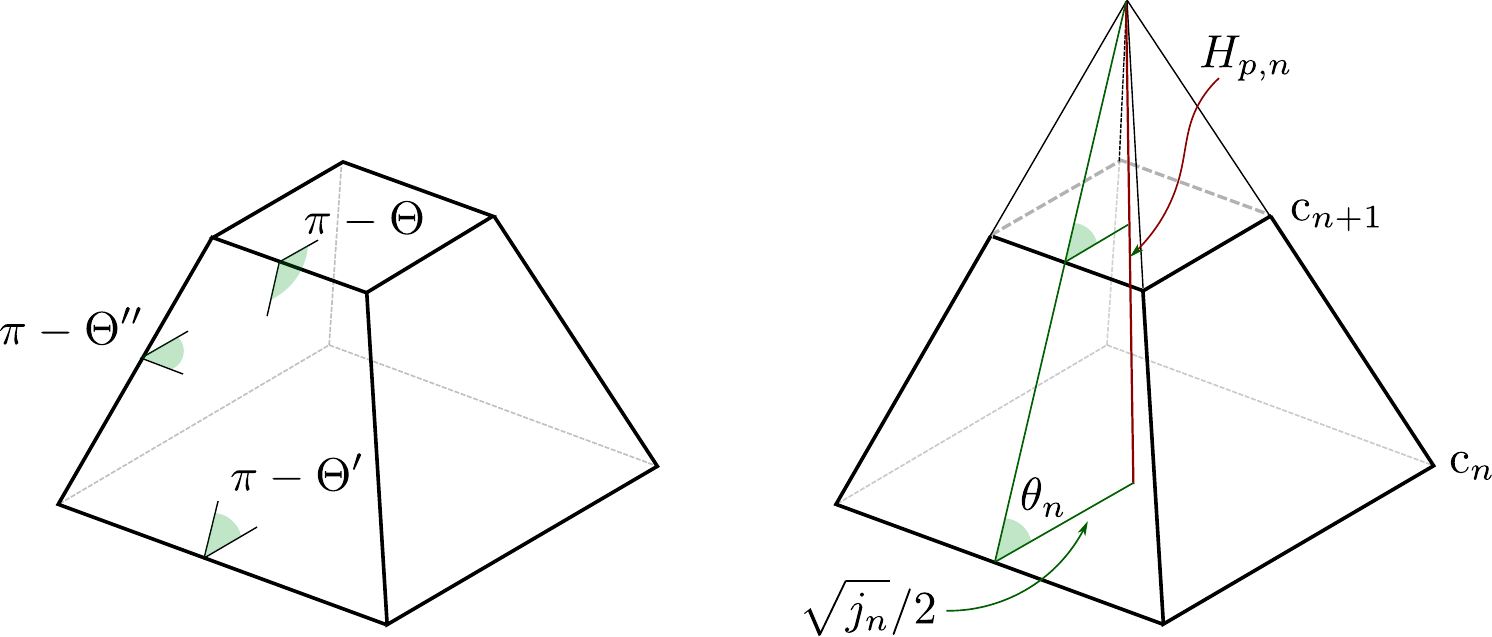}
	\caption{\small Three dimensional representation of a four dimensional hyperfrustum. The top and bottom square faces represent cubes. The side faces represent squared pyramidal frusta.}
	\label{fig:geometry}
\end{figure}

\subsection{Volume of a hyperfustum}
For the following analysis we refer to figure \ref{fig:line}.
The volume $V_n$ of the hyperfrustum $F_n$ that appears in equation \eqref{volume} can be computed as the difference of the volumes of two  four dimensional pyramids with base cubes $\mathrm{c}_n$ and $\mathrm{c}_{n+1}$. A comparison with the three dimensional representation in figure \ref{fig:geometry} on the right may be helpful to get an intuitive understanding.
The volume of the four dimensional pyramids with base cubes $\mathrm{c}_n$ and $\mathrm{c}_{n+1}$ are
\begin{equation}
V_{p,n}=\frac{1}{4}H_{p,n} j_n^{3/2}, \qquad \qquad V_{p,n+1}=\frac{1}{4}H_{p,n+1} j_{n+1}^{3/2}.
\end{equation}
being $H_{p,n}$ and $H_{p,n+1}$ the heights of the pyramids. These have to be determined in order to ensure that the `slope' of the hyperpiramidal sides is the same as for the hyperfrustum. Their values are
\begin{equation}
H_{p,n}= \frac{1}{2}\sqrt{j_n} \tan \theta_n, \qquad \qquad H_{p,n+1}= \frac{1}{2}\sqrt{j_{n+1}} \tan \theta_n,
\end{equation}
and they are constrained so that their difference gives the height of the hyperfrustum \eqref{heighthyperfr}. Finally, the four dimensional volume of the hyperfrustum is given by
\begin{equation}
V_n= V_{p,n} - V_{p,n+1}= \frac{1}{8}(j_n^2-j_{n+1}^2)\tan \theta_n.
\end{equation}
Using equation \eqref{thetangle} for the angle $\theta_n$, one can easily find the expression \eqref{volume} of the volume of the hyperfrustum in terms of the spin variables.

}


\bibliography{Bibliography9April}
\bibliographystyle{ieeetr}

\end{document}